\documentclass[12pt,preprint]{aastex}           

\usepackage{graphicx}                   
\input{psfig.sty}                       
\usepackage{rotating}
\usepackage{longtable}
\usepackage{url,hyperref,color}

\shorttitle{Pulsational time evolution study of AN Lyncis}
\shortauthors{Zhou et al. }

\begin{document}

\title{A Pulsational Time-evolution Study for the $\delta$ Scuti Star AN Lyncis
}

\author{A.-Y. Zhou\altaffilmark{}}
\affil{National Astronomical Observatories, Chinese Academy of Sciences,
      20A Datun Road, 100012 Beijing, China.}
\email{aiying@nao.cas.cn}

\author{ Eric G. Hintz, Jeremy N. Schoonmaker\altaffilmark{} }
\affil{  Department of Physics and Astronomy, Brigham Young University, N283 ESC, Provo, UT 84602, USA}
\email{hintz@physics.byu.edu}

\author{ Eloy Rodr\'{\i}guez, Victor Costa and M. J. Lopez-Gonzalez\altaffilmark{} }
\affil{ Instituto de Astrof\'{\i}sica de Andaluc\'{\i}a, CSIC,
      P.O. Box 3004, E-18080 Granada, Spain }
\email{eloy@iaa.es}

\author{ Horace A. Smith and Nathan Sanders\altaffilmark{} }
\affil{  Department of Physics and Astronomy, Michigan State University, East Lansing, MI 48824-2320, USA}
\email{smith@pa.msu.edu}

\author{ Gerold Monninger\altaffilmark{} }
\affil{  Bundesdeutsche Arbeitsgemeinschaft f\"{u}r Ver\"{a}nderliche Sterne e.V. (BAV),
Munsterdamm 90, DE-12169 Berlin, Germany }
\email{gerold.monninger@online.de}

\and

\author{ Lienhard Pagel\altaffilmark{} }
\affil{ Fakult\"{a}t f\"{u}r Informatik und Elektrotechnik,
Institut f\"{u}r Ger\"{a}tesysteme und Schaltungstechnik, Universit\"{a}t Rostock,
Albert Einstein Str. 2, 18051 Rostock, Germany }
\email{lienhard.pagel@uni-rostock.de}


\begin{abstract}
This paper presents a large amount of observations for
the $\delta$ Scuti star AN Lyncis carried out in 2001--2012.
The extensive observations include two tri-continent campaigns coordinated
in 2002 and 2011, respectively,
and several single-site contributions throughout the period.
The data in total have more than 104100 raw CCD frames and photoelectric records, which
consist of 165 nights (about 816 hours) spanning over 3778 days.
The final reduced light curves have
more than 26500 data points (including those 3462 unpublished BYU data),
from which we determined 306 new timings of maximum light.
A time-dependent behaviour study based on all available data indicates
cyclic amplitude variability as well as period change [for the main periodicity].
Orbital sinusoid fittings to $(O-C)$ residuals and pulsation amplitudes
may account for their variations being caused by the light-time effect
of AN Lyn in a binary system.
The orbital period is about 26--30 years.
Current results support the binarity of AN Lyn, first suspected by Zhou (2002).
We further show the detailed time evolution structure of the pulsation of AN Lyn
as function of both time and period through wavelet analyses.
\end{abstract}


%
%
\section{INTRODUCTION}
AN Lyncis (=BD +43$^{\circ}$1894, $\alpha_{2000}=09^{\rm h}14^{\rm m}28^{\rm s}.69$,
$\delta_{2000} = +42^{\circ}46'38^{\prime\prime}.2$ [Simbad: FK5],
$\langle V\rangle\sim$10\fm7, $\Delta V\sim$0\fm18, A7IV-V) has been
confirmed to be a medium-amplitude multiperiodic
$\delta$ Scuti variable of mixed radial and nonradial pulsations.
The current frequency solution for the pulsation of the star includes two
independent frequencies $f_1$=10.1756 and $f_2$=18.1310 cycle~d$^{-1}$ along with
the harmonics 2$f_1$ and their interaction terms $f_1$+$f_2$=28.3066 and
2$f_1$+$f_2$=38.4822~\citep{rodr97a,rodr97b,zhou02}.
Our previous work shows that the star seemed to present cyclic amplitude variations
during the period from 1980s to 2001 (\citealt{zhou02}).
\citet{hintz05} concluded that AN Lyn is a member of a binary system
in terms of their radial velocity and $(O-C)$ data.
These authors also showed a mysterious phenomenon that the amplitude of pulsation
is changing in sync with the orbital motion (of period about 22 yr) derived from
their $(O-C)$ diagram. A recent study by \citet{liq10} agrees with this synchronized cyclic change and
accounts for it by solar-type magnetic activity in the star.

Previous work has posed further interesting questions about the amplitude variability and
period change of AN Lyn.
The fact is that neither a linear nor quadratic fit is good for the $(O-C)$ graph,
while the amplitude variation seems sinusoidal.
To re-check the pulsation frequencies, the variation in the trend of pulsation amplitude,
as well as the suspected binarity,
we twice undertook tri-continent campaigns on the star in the spring of 2002 and 2011, respectively.
Besides the campaigns, extensive data from single-site observations were collected in 2001--2012.
Based on these data, we conduct a comprehensive analysis
for the star's binarity and pulsation.
This paper reports the observational data along with a time evolution study of
the primary pulsation frequency and its amplitude variability.
In a forthcoming paper, we will focus on an exhaustive investigation
for pulsation components and modes identification with the color photometry data.

\section{OBSERVATIONS}
\label{sect:Obs}

\subsection{The 2002 tri-site campaign}
\label{2002camp}
The tri-site campaign observation dates were set in 2002 February.
Due to bad weather conditions and instrument problem,
one site failed to observe during this interval, but a few data were collected after
the coordinated time. Details were given below.

In Xinglong (China), between 2002 February 1 and 28, and between December 27 and 30,
differential photometry was carried out with
a three-channel high-speed photoelectric photometer \citep{jiang98} mounted on
the 85-cm telescope at the Xinglong Station of the National Astronomical Observatories
of the Chinese Academy of Sciences (NAOC, formerly known as BAO), China.
The detector is a copy of that previously used
by the Whole Earth Telescope campaign \citep[WET;][]{nat90}.
The variable star, comparison star and sky background were simultaneously exposed in
continuous 10-s intervals through the standard Johnson $V$ filter.
Differential magnitudes of AN Lyn were measured relative to
the comparison star GSC 02990-00019 ($\alpha_{2000}=09^{\rm h}14^{\rm m}34^{\rm s}.77$,
$\delta_{2000}=42^{\circ}41'02^{\prime\prime}.7$, {\em V}=10\fm6),
which is actually two superimposed objects with
a faint visual companion (2MASS J09143473+4241023, {\em B}=11\fm9).
But it was observed as a nonvariable within the typical observational accuracy
of about \hbox{$\pm$0.005\,mag}.
The nightly light curves data have been merged into either 60-s or 120-s bins
in terms of their quality.

In Granada (Spain), simultaneous Str\"{o}mgren $uvby$ photometric observations of AN Lyn were
carried out in 2002 February using
the six-channel $uvby\beta$ spectrograph photoelectric photometer
attached to the 90-cm telescope at the Sierra Nevada Observatory\citep[SNO,][]{rodr97a}.
With this instrument, in each integration time one target is exposed on four
({\em uvby}) bands of the Str\"{o}mgren photometric system
or two bands (narrow ``n'' and wide ``w'' bands of the Crawford H$_{\beta}$ system)
simultaneously.
In these observations, only the mode \emph{uvby} was used.
Sequential observations were usually performed in cycles consisting of
``sky, C1, C2, Var''.
HD 78512 and HD 78572 were used as comparison (C1) and check (C2) stars,
similar to \citet{rodr97a,rodr97b}.
A duty cycle usually took five minutes.

In East Lansing, Michigan (USA), CCD photometry was done on four nights
between 2002 May 10 and June 1, using an Apogee AP47p CCD camera and the $V$ filter
on the 24-inch Boller \& Chivens telescope of the Michigan State University (MSU) campus observatory.
Differential magnitudes were measured relative to GSC 02990-00019.
A realistic estimate of the photometric uncertainty of a typical observation is
about \hbox{$\pm$0.015\,mag}~\citep[refers to][]{lac01}.
Post-reduction revision of aperture photometry using IRAF (see next section)
improved the accuracy to about \hbox{$\pm$0.01\,mag}.

\subsection{Single-site contributions in 2001--2012}
CCD photometry on several telescopes was performed individually at several sites
during 2001 and 2012 using different instrumental configurations.
They are summarized as follows.

\begin{itemize}
  \item In 2006, either a Princeton Instruments RS1340b CCD camera or
an Apogee Alta U47+ 1024$\times$1024 CCD was used on
the 40-cm telescope at the Baker Observatory of Missouri State University,
Springfield, Missouri, USA.
The images were binned in 2$\times$2 pixels.
We secured several hours of data covering four maxima on three nights
with Johnson $V$ filter.

  \item In 2007, 2009 and 2010, observations were acquired with either
an Apogee AP7P CCD camera sized 512$\times$512 or
a Princeton Instruments MicroMAX:1024BFT CCD \cite[MiCPhot,][]{zhou09} mounted on
the 85-cm telescope at the Xinglong Station of NAOC, China.
Observations were all made in Johnson $V$ band.
We acquired 43, 2 and 10 nights data in 2007, 2009 and 2010, respectively.

  \item In 2008, a few Johnson $V$ data covering four maxima were
  obtained on three nights by the BFOSC system \citep{huang05} on the 2.16-m telescope of NAOC, China.

  \item In 2004--2010, a number of standard single $V$ or multicolor $BVI$ data were
  collected with either
  an SBIG ST-10XME CCD camera on the 0.4-m David Derrick Telescope, at the Orson Pratt Observatory,
  which is located in the center of the Brigham Young University (BYU) campus,
  or an SBIG STL-1001 CCD on the 0.51-m telescope at the BYU West Mountain Observatory, Provo, Utah, USA.
  Image reduction and photometry were done by the observers.
  The observations yielded an error per observation on the order of 0.004\,mag~\citep[refers to][]{hintz09}.

  \item
In 2001--2012, a total of 1841 differential measurements related to GSC 02990-00019 were
collected on 9 nights (38.5 hours, covering 14 maxima) using a 14-inch Cassegrain telescope
at f/9 or f/6 by Gerold Monninger (BAV) at a private observatory in Gemmingen (Germany).
The CCD photometry observations were started with the observatory's SBIG ST-6 CCD camera
using an IR cut-off filter (2001) and V filter (2002--2011).
The effective field of view of the CCD photometric system is about 9.5'$\times$7.2'
and the size of each pixel is 1.5$^{\prime\prime}\times1.8^{\prime\prime}$.
Further time-series CCD photometry was obtained using a CCD SBIG ST-10XME and
V filter in 2011--2012.
The CCD was configured in a 3$\times$3 binning mode resulting in an angular resolution of
2.0$^{\prime\prime}$/pixel (the field of view is 24'$\times$16').
The CCD images were reduced with standard procedures in Mira AP.
The flatfield correction utilized sky-flat images taken during the morning twilight.
Aperture photometry was also performed in Mira AP and differential magnitudes were calculated.

  \item From 2012 March 15 to May 27 (UT), about 26 hours of observation covering 14 maxima on
13 nights were obtained with the 10-inch Schmidt-Newton telescope (f=1.0\,m) at Klockenhagen, Germany.
An Artemis 4021 CCD camera binned in \hbox{2$\times$2} was used.
This camera has a pixel size of \hbox{7.4\,$\mu$m} and chip size of \hbox{15.16$\times$15.16\,mm$^2$}.
It uses the Kodak KAI4021 sensor which has 2048$\times$2048 pixels.
On 2012 April 17, observation was first started at 20:37:20 UT,
then the parallel observation was launched at 20:52:14 UT
on the 18-inch Newton telescope (f=2.0\,m) equipped with the QHY8 CCD camera,
which has active pixels of 3021$\times$2016 (sized \hbox{17.64$\times$25.1\,mm$^2$}
with a pixel size of \hbox{7.8\,$\mu$m}).
This camera takes the RGB BAYER film on CCD as color method.
The images was taken in the `G' color.
The CCD chip was thermoelectrically cooled to \hbox{$-20^{\circ}$C}.
In addition, one maximum was obtained on 2010 April 2 without filter by the 10-inch telescope.
The observations were done with the standard $V$ filter except those specified.
The field of view includes the specified comparison stars
GSC 02990-00019, 00115, 00221, 00575 and 00549, which were used in calibration.
\end{itemize}

\subsection{The 2011 tri-continent campaign}
To finalize our observing plan on AN Lyn, observations
were ended up by the second time coordinated tri-continent campaign
composed of four telescopes at SNO in Spain, NAOC in China, and MSU and BYU in America
during the first half of 2011 February.
This time, as we wished for mode identification of pulsation frequencies,
we asked each participant to observe the target with Johnson $BVI$ filters.
The observations were summarized below.

\begin{itemize}
  \item BYU run: In 2011 Feb.1--14, 192 frames were obtained in each of $VI$ filters on 5 nights,
while 134 frames for $B$ filter on 3 nights. The instrument is an SBIG STL-1001 CCD camera
attached to the 0.4-m David Derrick telescope,
at the Orson Pratt Observatory of BYU, Provo, Utah, USA.
  \item MSU run: In 2011 Feb.1--14, 301 frames in each of $BVI$ filters were collected
  on two nights with the same telescope described in Sect.~\ref{2002camp}.
  The CCD camera used was an Apogee Alta U47.
  \item NAOC run: In 2011 Feb.1--14, observations were carried out with the MiCPhot
  on the 85-cm telescope at NAOC, China.
Sequential observations were made in Johnson $BVI$ filters.
We acquired 3112 stellar frames in each filter on 8 nights,
totaling to 65.8 hours of multi-color data.
  \item SNO run: During the coordinated campaign days 2011 Feb.1--14,
  due to instrumental problems no observation was secured.
After the campaign time,
a few hours of CCD imaging were obtained during the nights of
March 29 and 30 using a Princeton Instruments 2K$\times$2K imaging camera of
model Roper Scientific VersArray 2048B on the 1.50-m telescope at SNO.
This camera is based on a high quantum efficiency back-illuminated CCD chip
with enhanced response in the ultraviolet, the Marconi-EEV CCD42\_40 chip.
On April 12, 13 and 14, a total of 286 stellar images were collected
using a new 4K$\times$4K CCD camera
of SBIG STX-16803 
mounted on the 90-cm telescope.

The field-of-view of a frame obtained on these two telescopes are
about 7.9$\times$7.9 and 17$\times$17\,arcmin$^2$, respectively.
Binning of 2$\times$2 was applied throughout the run
in order to decrease the readout time and improve the signal-to-noise ratio as well.

The images were obtained in a sequential way for either
the three Johnson filters \emph{B, V} and \emph{I} or
the three custom filters $b$ (Blue), $g$ (Green) and $r$ (Red).
The typical integration times were of about 20, 15 and 10 s for
filters \emph{B}, \emph{V} and \emph{I} during the first night;
and of 11, 8 and 6 s during the second night.
While the typical integration times were of 30 s for the scientific frames in each
custom filter.
These three custom filters are centred at the following wavelengths:
$b$ filter at 4500\AA~(full width at half-maximum about 1250\AA),
$g$ filter at 5400\AA~(FWHM$\sim$860\AA),and
$r$ filter at 6400\AA~(FWHM$\sim$1050\AA).
Hence, these $bgr$ filters \citep{stra92} resemble the Johnson-Cousins filters $B$ (centred at 4350\AA),
$V$ (centred at 5470\AA; or Str\"{o}mgren $y$ filter centered at 5470\AA) and
$R$ (centred at 6580\AA).
\end{itemize}

\begin{figure*}[tp!]     
\vspace{-9mm}
\centering
\includegraphics[width=0.9\textwidth,height=160mm,angle=-90,clip=true]{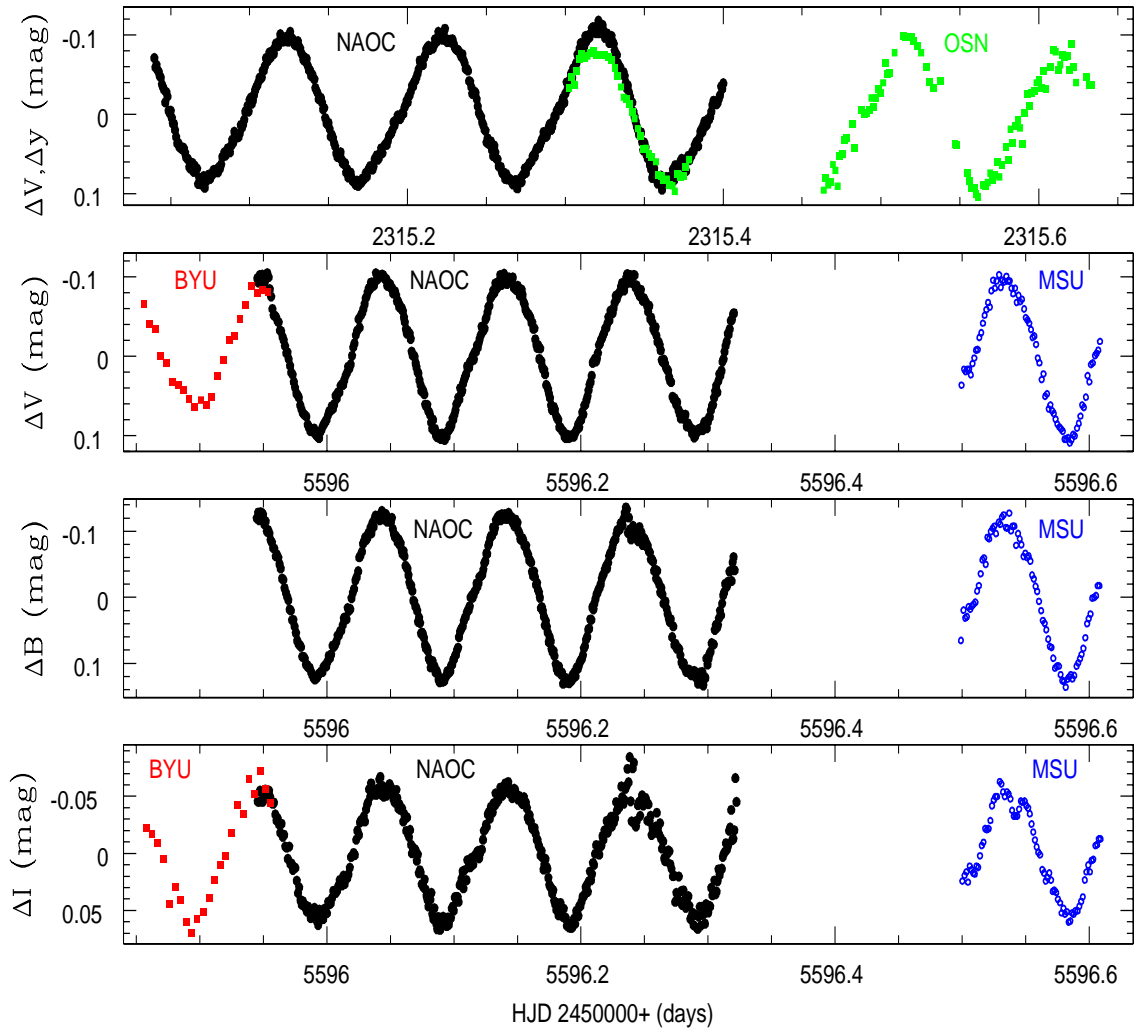}
   \caption{ The representative campaign light curves acquired in 2002 and 2011.
Top panel: $V$ and $y$ data on 2002 February 9;
Second to last panels: $VBI$ data on 2011 February 3.  }
\label{Fig:LC-camp}
\end{figure*}

\begin{figure}[tpb]
\vspace{-8mm}
\centering
\includegraphics[width=120mm,height=90mm,angle=-0,clip=true]{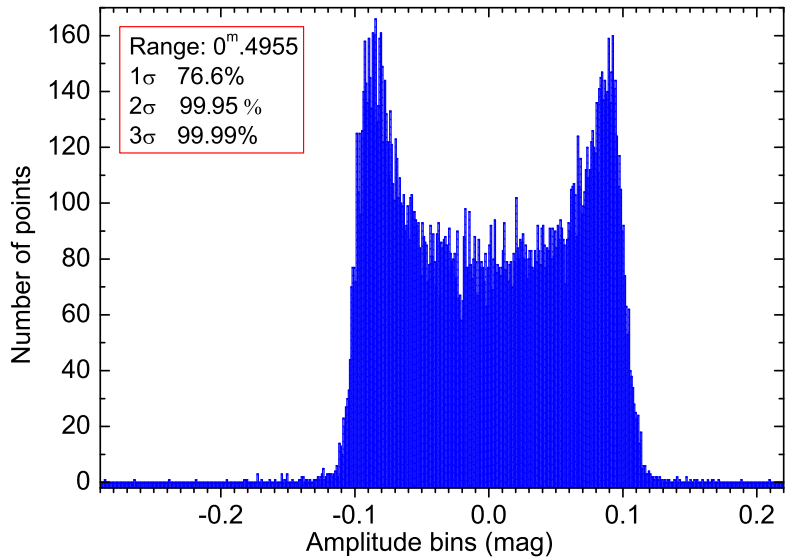}
\caption{A statistics of the 2001--2012 light curve data of AN Lyn. `$1\sigma$' valued 0.06626\,mag. }
\label{Fig:2002-2012LC-stat}
\end{figure}

The participating telescopes and instrumentation are summarized in Table~\ref{Tab:obs-telescope}.
Details on all the observations are tabulated in Table~\ref{Tab:Log}, which is
available online in electronic version only\footnote{
Table 2: \url{http://journal-web/ANLyn-table2.tex}}.
In sum, we got about 816 hours of useful data with more than 104100+ raw CCD frames and photoelectric records,
on a total of 165 nights spanning over 3778 days since 2001.
Representative campaign light curves are given in Fig.~\ref{Fig:LC-camp},
while all the light curves are provided in the online electronic version
as Figs.~\ref{Fig:LC2002NAOC}--\ref{Fig:LC2012Pagel}.
Statistics of the whole data are presented in Fig.~\ref{Fig:2002-2012LC-stat}.
The time-series data are available for downloading in three text table
files\footnote{
 2002 $uvby$ light curve data: \url{http://journal-web/ANLyn2002-uvby.txt},\\
 2002--2012 $V$ light curve data: \url{http://journal-web/ANLyn2002-2012-V.txt},\\
 2011 $BVI$ light curve data: \url{http://journal-web/ANLyn2011-BVI.txt}}.
Times are always given in Heliocentric Julian Date (HJD).

\placefigure{Fig:LC-camp}      
%

\placefigure{Fig:2002-2012LC-stat}          
%

%
\begin{table}
   \caption[ ]{Participating sites, telescopes and instrumentation.   }
   \label{Tab:obs-telescope}
   \begin{center}\begin{tabular}{llrlc}
   \hline\noalign{\smallskip}
Observatory &    Location     &Telescope& CCD Detector or PM & Filters \\ 
\hline\noalign{\smallskip}
Baker   & Springfield, MO, USA&   40-cm & Apogee Alta U47+     & $V$     \\
            &                 &         & PI RS1340b    & $V$     \\
BAV      & Gemmingen, Germany &   35-cm & SBIG ST-6     & $V$     \\
            &                 &         & SBIG ST-10XME & $V$     \\
BYU         & WMO,Provo,UT,USA&   51-cm & SBIG STL-1001 & $V$     \\
            & OPO,Provo,UT,USA&   40-cm & SBIG ST-10XME & $BVI$   \\
Klock  & Klockenhagen, Germany&   25-cm & Artemis 4021  & $V$     \\
            &                 &   46-cm & QHY8          & G of RGB\\
MSU    & East Lansing, MI, USA&   60-cm & Apogee AP47p  & $V$     \\
            &                 &         & Apogee Alta U47      & $V$     \\
NAOC        & Xinglong, China &   85-cm & 3-ch photomultiplier&$V$     \\ 
            &                 &         & Apogee AP7P   & $V$     \\ 
            &                 &         & PI MicroMAX:1024BFT&$V$ \\ 
NAOC        & Xinglong, China &  216-cm & BFOSC system  & $V$     \\ 
SNO         & Granada, Spain  &   90-cm & 6-ch photomultiplier&$uvby$  \\ 
            &                 &         & SBIG STX-16803& $bgr$ \\ 
SNO         & Granada, Spain  &  150-cm & PI VersArray 2048B& $BVI$ \\ 
   \hline\noalign{\smallskip}
   \end{tabular}
   \end{center}
\end{table}

\section{DATA REDUCTION}
\label{sect:Reduction}

\subsection{Multi-channel photoelectric photometry}
In 2002, photometry performed at SNO was simultaneous four ($uvby$) colors on multiple channels
each integration time for an object.
The SNO observational data were reduced by comparison with \hbox{HD 78512} (C1) and \hbox{HD 78572} (C2),
and calibrated by taking use of a number of regularly observed photometric standard stars.
Detailed reduction procedures refer to \citet{rodr97a}.
The standard deviations of (C2--C1) are 0.0081, 0.0048, 0.0046, 0.0057, 0.0053, 0.0055, 0.0108\,mag
for $u, v, b, y, (b-y), m_1, c_1$, respectively.
The observations carried out at NAOC recorded three targets once exposure
on multiple channels simultaneously.
The NAOC 3-channel data were reduced quite directly by first calculating sky response coefficients
among the three channels, and then producing differential magnitudes from
sky-subtracted counts for variable and comparison channels.
Due to short exposure (10 seconds) of the three-channel raw records, they were binned to
produce equal integration of 60 or 120 seconds per point for reducing scatter as well as
balancing weights in the final analysis.
For the data of good quality, over-binning is avoided to prevent the details of
light variations from being smoothed out. This policy is kept throughout
all the reductions for both photoelectric and CCD data.

\subsection{CCD photometry}
Aperture photometry was employed to extract magnitudes from the calibrated frames.
The procedures of CCD images reduction included
bias and dark subtraction and flatfield correction,
as well as the aperture-optimized photometry.
The CCD photometry reductions were implemented with IRAF, as outlined in \citet{zhou06}.
The photometry program measures the target and several reference stars in six apertures
and builds differential magnitudes for the target relative to
different numbers of reference stars.
Which aperture and how many reference stars are finally chosen depends on whether or not
the resulting differential magnitudes are of the least scatter.
The photometry apertures may vary from frame to frame as they are optimized based on
the average full width at half maximum (FWHM),
which is determined via three well-exposed stars in each frame.
The selected reference stars in the field of AN Lyn were detected as non-variables
within the observational error of 0\fm005,
a derived typical accuracy representing the standard deviation of
the differential magnitudes between any two reference stars.
Then we did ensemble differential photometry for AN Lyn.
Differential magnitudes for the target were finally yielded either by using
a single comparison star (for some frames obtained at MSU)
or an ensemble comparison sequence comprised of two to five reference stars.

\subsection{Light curves alignment}
\label{sect:align}
\begin{figure*}[tp]     
\vspace{-14mm}
\centering
\includegraphics[width=0.9\textwidth,height=115mm,angle=-0,clip=true]{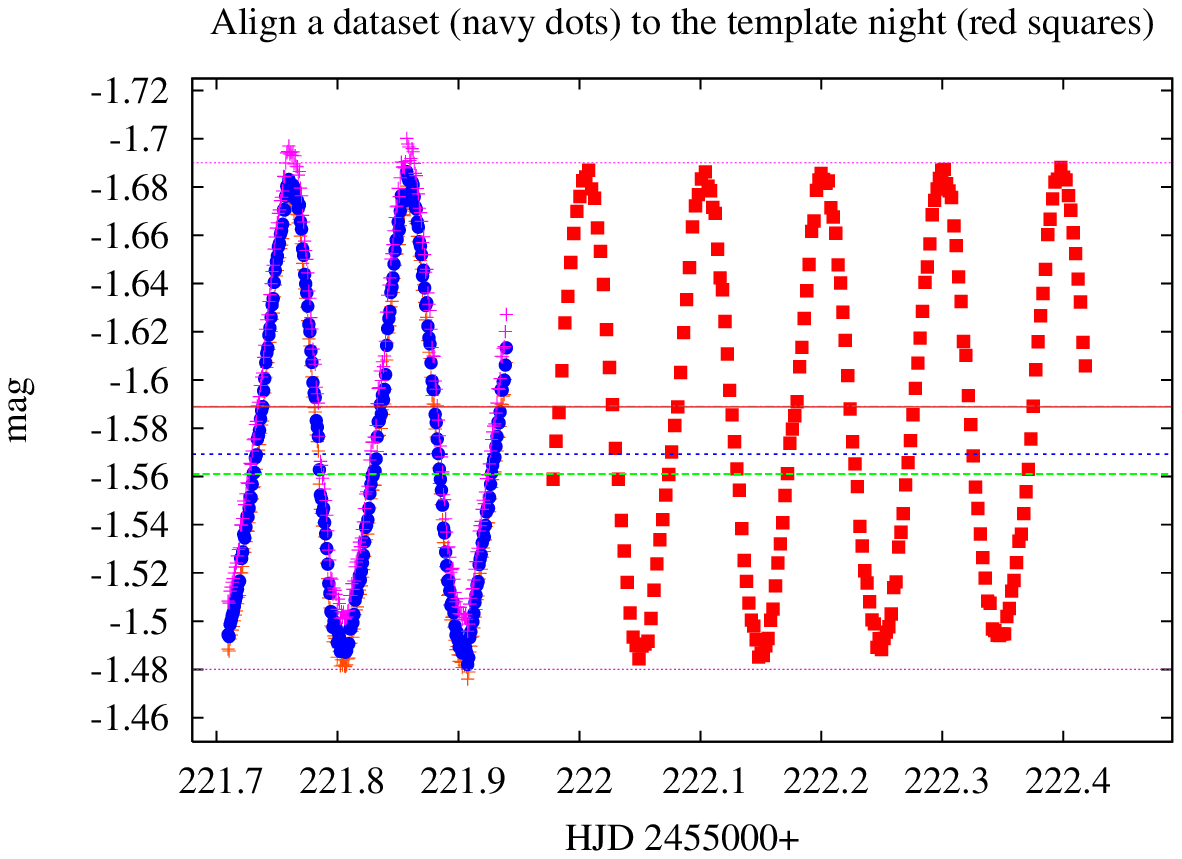}
\vspace{-1mm}
\caption{ Illustration of light curves alignment:
   the aligned guest LC (navy dots on left) is resulted from applying
   a shift of 0.006 mag in y-axis to its raw values.
   This shift is not equal to the differences in mean between the unaligned guest LC (in brown crosses) and
   the template LC (red squares on right), 1.58909$-$1.569245=0.019845,
   or the differences in median, 1.5888$-$1.5610=0.0278\,mag.
   The magenta crosses show the movement by either of the differences.
   Magenta lines: range of the template;
   three middle lines from lower to higher:
   median of guest in green, mean of guest in blue, mean/median of template in red;
   The guest LC was shifted in x-axis for display.}
   \label{Fig:LC-align}
\end{figure*}

As observations come from different configurations of telescopes and photometers,
and from multiple observing seasons, and, moreover,
differential magnitudes are derived from non-uniform comparison stars,
nightly light curves are not uniformly on the same mean level.
However, we are aware that the true mean value for the overall light curves,
which should be subtracted in Fourier analysis
\footnote{Inconsistent mean levels
could cause noise and spurious peaks in the low frequency domain in a Fourier spectrum},
may be incorrectly calculated due to outlying measurements and non-uniform nightly means.
The arithmetic mean is defined as the average value of a set of data sampled in full or multiple cycles
is usually reliable to be subtracted for a night,
but it is not always a good idea to do so
for those light curves covering incomplete cycles.
We are sure that the real mean value of a nightly light curves sampled in incomplete cycles
will be biased to either side of the zero mean level of a normal sine curve.
The median of a set of data may not be affected by several outliers,
but it could be significantly biased due to incompletely sampled cycles in practice.
In short, neither the mean nor the median of a night's light curve can be simply subtracted
for use in Fourier analysis.
In take account of this, we take a night's light curves of good quality
with the best cycle coverage in its run or season as `template', then
we compare other nights light curves (call `guest') with this `template' to align them.
Indeed, for all data we plot a `guest' night light curves along with the `template' light curve
for comparison.
We look at the graph and inspect how much the `guest' light curve deviates from
the `template',
then we shift the `guest' light curves vertically by a trial value.
The graph is immediately redrawn to display the shifted `guest' light curves.
We repeat this procedure until a satisfactory alignment of two is reached,
that is, the `guest' light curves are properly positioned relative to the `template'
in the graph window.
Much attention has been paid to the data sampled over incomplete cycles,
those short light curve pieces that cover less than a full cycle.
We make sure that the `guest' light curves are well placed in the right position of its phase
according to the `template' data.
The light curves alignment was done night by night and season by season.
Lastly, we constructed seasonal data sets for use in the pulsational analysis.
Prior to put seasonal data sets together we subtract their seasonal means.
This aligning procedure should reduce low-frequency noise in Fourier spectra and
then improve the zero point of nonlinear sine-wave fitting.
Figure~\ref{Fig:LC-align} 
illustrates the aligning method.
%
\placefigure{Fig:LC-camp}      

\subsection{Measuring timings of maximum light}
With the light-curve data we have determined a total of 306 new times of maximum light
following the method illustrated in fig.4 of \citet{zhou03},
i.e. applying a 2nd--6th order polynomial fit to a proper portion of the light curves
around a light maximum and then deriving the extremum.
The error for maximum determination is normally about 0\fd00045 or less.
The new maxima are reported in Table~\ref{Tab:max306} in the electronic version online.
A list of all the 541 maxima analyzed in this paper is also provided online
in Table~\ref{Tab:max541}.

\section{DATA ANALYSIS}
\label{sect:Analysis}
Stellar pulsation is usually unstable, so we wonder how much the pulsation varies over time,
and in what a manner the pulsation periods and even amplitude evolve with time.
In this section, we analyze the light variations of AN Lyn
with a special interest in examining its temporal evolution behaviour.

\subsection{Pulsation timescales and variability}
Autocorrelation analysis is a simple method for looking for cyclic behavior in variable star data.
This method is able to detect characteristic timescales averaged over all the data.
It is excellent for stars with amplitude variations and/or transient periods.
We used {\sc astrolab}, a program developed by Prof. John Percy and his students
at the University of Toronto \citep{percy01},
which allows one to perform self-correlation analysis for time-series data.
In this method, for all pairs of measurements,
the difference in magnitude ($\Delta m$ in a number of bins) and
the difference in time (time lag, $\tau$)
to some upper limit are calculated.
This limit was set to be a few times greater than the expected timescales
but much shorter than the total time span of the data.

\begin{figure}[t!]
  \vspace{0mm}
\centering
\includegraphics[width=0.75\textwidth,angle=-90,clip=true]{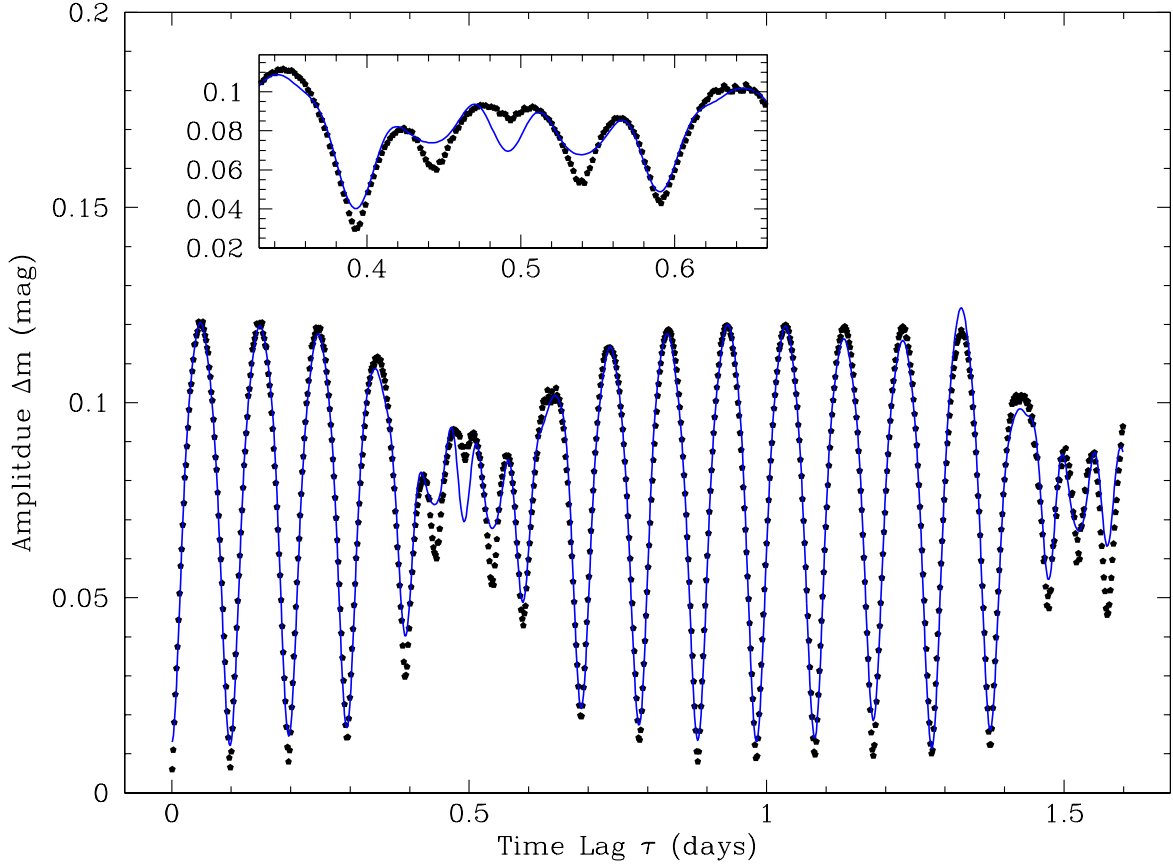}
\caption[]{Autocorrelation presentation of the amplitude and period variability
of AN Lyn for the 2007--2011 data set.
The solid line is a 10-frequency fit.
Deviation between the fit and $\Delta m$ around 0.5-day time lag is
clear in the enlarged insert window.}
\label{Fig:autocorr}
\end{figure}

\placefigure{Fig:autocorr}
%

\begin{figure}[ht!]
\centering
\includegraphics[width=0.70\textwidth,angle=-90]{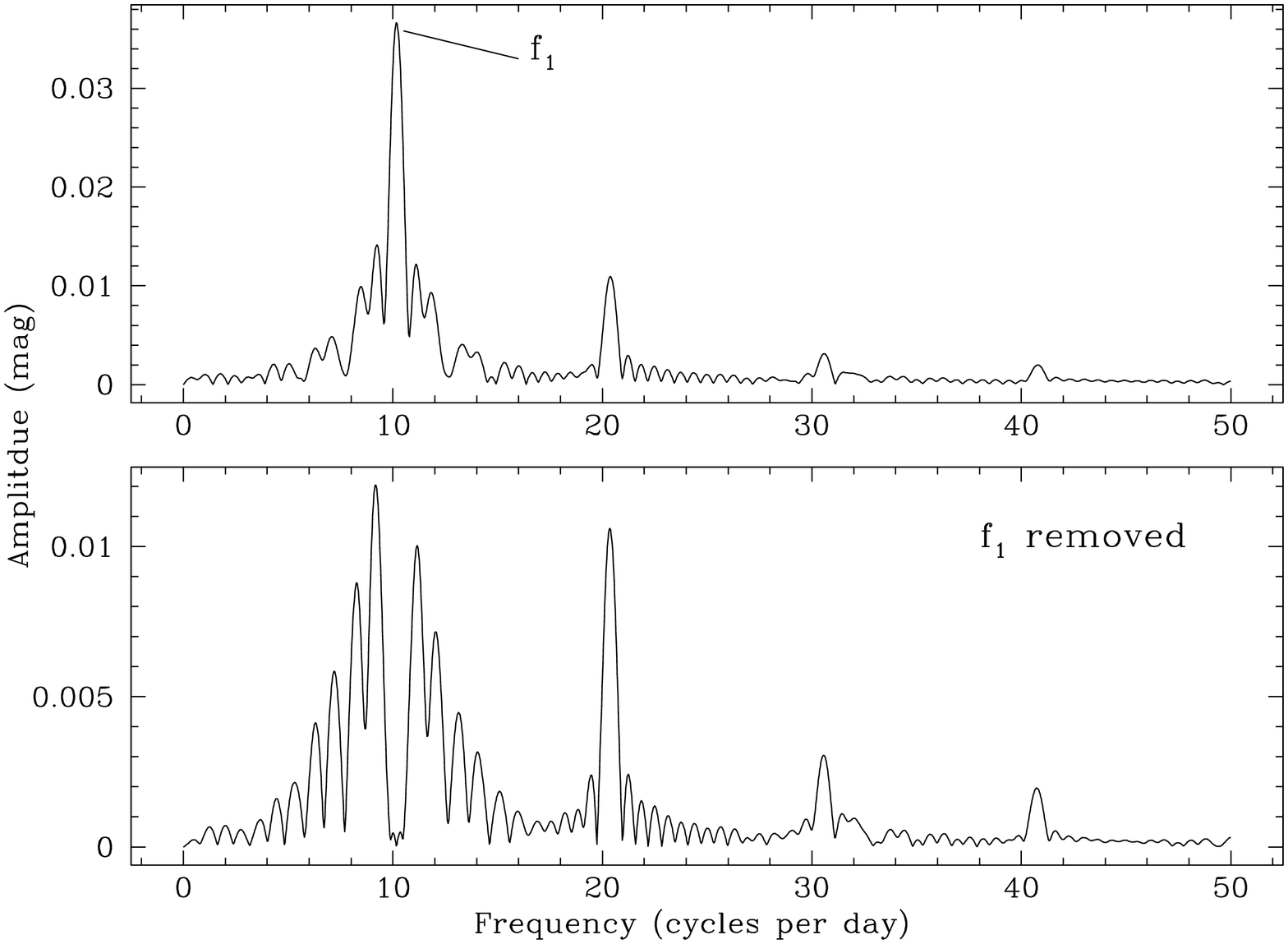}
\caption[]{Fourier spectrum of the autocorrelation results.}
\label{Fig:ACF-FT}
\end{figure}

\placefigure{Fig:ACF-FT}

In Fig.~\ref{Fig:autocorr},
we plot $\Delta m$ against $\tau$ for the data in 2007--2011.
The average $\Delta m$ is a minimum at multiples of $\tau$.
Each minimum can be used to estimate $\tau$ ---
here the characteristic cyclic light variation timescale, i.e. pulsation period.
From the figure we can see that the minima
occurs increasingly prior to integer multiple locations of 0.098 days as the time lag increases.
This means that the primary period (0.098 days) is changing
or the light variations are multiperiodic.
Secondly, the height of the maxima is a measure of the average amplitude of variability.
Thus the maxima indicate clear amplitude variations.
The height of the minima in Fig.~\ref{Fig:autocorr} are actually determined
by the average error of the magnitudes and by the degree of irregularity in the pulsation.
If the variability were perfectly mono-periodic and the magnitudes had no error,
then the minima would fall to zero.
The minima are in the range of 0.005--0.012 mag.
Figure~\ref{Fig:autocorr} shows the change of the height of maxima and the minima positions,
i.e. cyclic variation of both amplitude and period.
The over-plotted solid line is a multi-frequency fit (of 10 terms), which still deviate from
the data. That means multiperiodic pulsation does not account for the autocorrelation results completely.
Unfortunately, self-correlation analysis is not so helpful in determining multiple periods.
The Fourier spectrum of the autocorrelation results (see Fig.~\ref{Fig:ACF-FT}) is, however,
a good presentation of the harmonics along with the modulation to the primary pulsation frequency.

\subsection{Amplitude of pulsation}
\begin{figure}[t]
   \vspace{2mm}
   \begin{center}
   \includegraphics[width=0.80\textwidth,height=90mm,angle=0,scale=1.0]{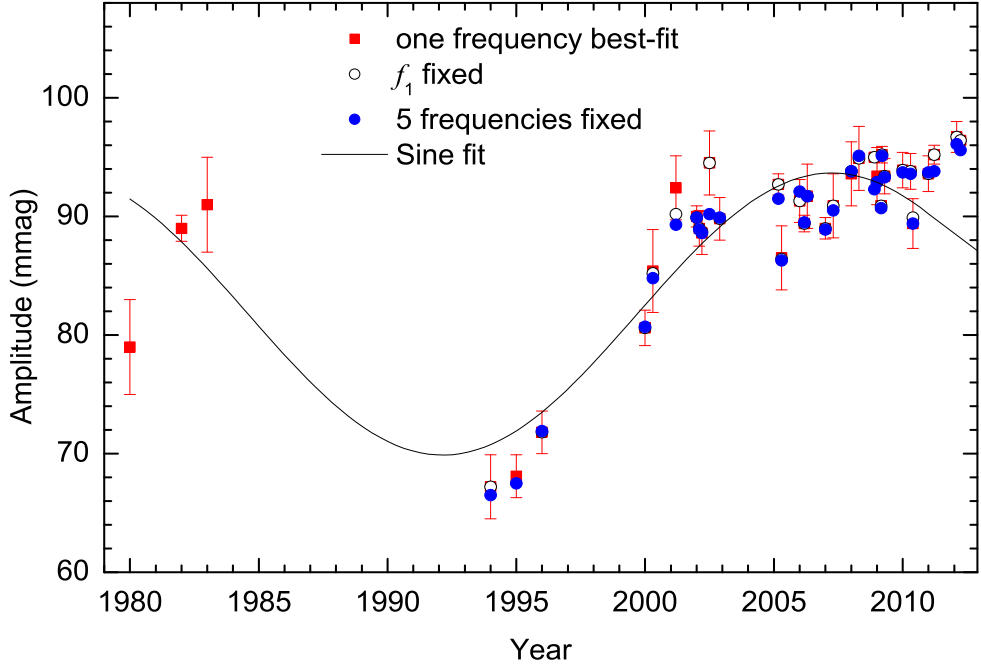}
\vspace{-2mm}
   \caption{Time evolution of the main pulsation amplitude of AN Lyn. }
\label{Fig:ampl}
\end{center}
\end{figure}

The amplitude of pulsation cannot be simply taken from peak-to-peak values of the light curves,
instead, it is often determined by Fourier analysis.
The Fourier amplitudes are `semi-amplitudes' of a sine-wave signal.
For all available light-curve data sets, we have calculated three kinds of
least-squares sinusoidal fits using the standard formula:
$ {\rm mag} (t) = Z + \sum_{i=1}^{n} A_i \sin [2\pi (f_i t + \phi_i)] $
via
(1)
best-fit allowing simultaneous improvement of
all the three Fourier parameters (frequency, amplitude, and phase) of a single frequency
with the trial value of $f_1$=10.1756\,d$^{-1}$;
(2) fitting with the main frequency $f_1$ fixed and allowing it to vary in amplitude and phase;
(3) fixing the five frequency terms
$f_1$=10.1756,
2$f_1$=20.3512,
$f_2$=18.1310,
$f_1$+$f_2$=28.3066 and
2$f_1$+$f_2$=38.4822\,d$^{-1}$
allowing amplitude and phase variations of these five terms.
Fittings were performed on each subset of light curves so as to determine
the mean seasonal pulsation amplitudes.
The amplitudes of $f_1$ determined in the three ways above are
listed in Table~\ref{Tab:ampl} and are drawn as a function of time
in Fig.~\ref{Fig:ampl}.
The error bars for the amplitudes of $f_1$ are the values
that were estimated high enough to conform with observational errors and data length.
Amplitudes for the 1980s (with error bars of 0\fm003--0\fm005) were adopted from the literature.
Among these three sets of values,
disagreement was obvious for the 2001 and 2002 MSU data, and the 1995 SNO data.
The reason could be attributed to fewer data points.
It is known that the Fourier amplitudes are averaged over a period of time
and they are obtained severally from different data sets
on the assumption that the frequency is constant during the course of data sampling.
We stress this point for a different viewpoint given by
wavelet transform in the next section.

\placefigure{Fig:ampl}

We tried a Fourier transform on all the amplitudes.
Then a nonlinear least-squares fit led us to the result
\begin{equation}
 \Delta V  = 81.7667 + 11.8899 \sin[\frac{2\pi}{30.21}(T-126.7224)]
\label{eq:ampl-sin}
\end{equation}
with a standard deviation of residuals of $\sigma$=1.979 (or reduced $\chi^2$ of 3.917).
The analytic standard errors for each of the four parameters above are 1.14, 1.5, 0.69 and 86.7, respectively.
The amplitude in the above fit is given in milli-magnitude and T is the time in years.
This fit suggests a cyclic amplitude variation with a period of 30.21$\pm$0.69 years
(see the solid line in Fig.~\ref{Fig:ampl}).
Note that the last points (from 2012 data) diverges from the sine curve.

\subsection{Long-term period variations }
In order to see the temporal-dependence of the main frequency during
the years with available data,
we applied the classic $O-C$ method to examine long-term period change.
The database of AN Lyn maxima has been recently updated by \citet{liq10},
who analyzed a number of 203 maxima including the 28 maxima reported by \cite{wils10}.
A note by \cite{hintz10} reported 18 maxima.
Together with our 306 new maxima and a few unpublished maxima from the BYU data archive,
we have a total of 541 maxima
at our disposal (see Tables~\ref{Tab:max306} and~\ref{Tab:max541},
which are available in the electronic version online).

To calculate $(O-C)$ residuals and their corresponding cycles (denoted by {\em E} below)
elapsed since an initial maximum epoch,
we first defined a trial ephemeris
\begin{equation}
 {\rm HJD_{max}} = 2452307.06608 + 0.0982743\times E
\label{eq:trial}
\end{equation}
after reviewing the ephemerides used in the literature
\citep{costa84,hintz05,wils10,liq10}.
The adopted period corresponds to $f_1$=10.1757\,d$^{-1}$ and
the initial epoch was taken from the light curves obtained on 2002 February 1.
The epoch was carefully determined by taking the mean of
three polynomial fittings of 3rd, 5th and 6th-order, which are applied to
different portions of the light curves around the maximum peak.
This precise initial epoch is expected to ensure that
the computed cycles for other maxima are not likely to be miscounted.
Then we applied a linear fit only to the 104 maxima collected in 2007
at NAOC --- a much consecutive set of data. Thus we refined eq.~(\ref{eq:trial}) to be~
\begin{equation}
{\rm HJD_{max}} = 2452307.077419 + 0.098274455\times E~ .
\label{eq:new-epoch}
\end{equation}
%

\begin{figure}[t!]
\vspace{-0mm}
\centering
\includegraphics[width=145mm,height=160mm,angle=-0,clip=true]{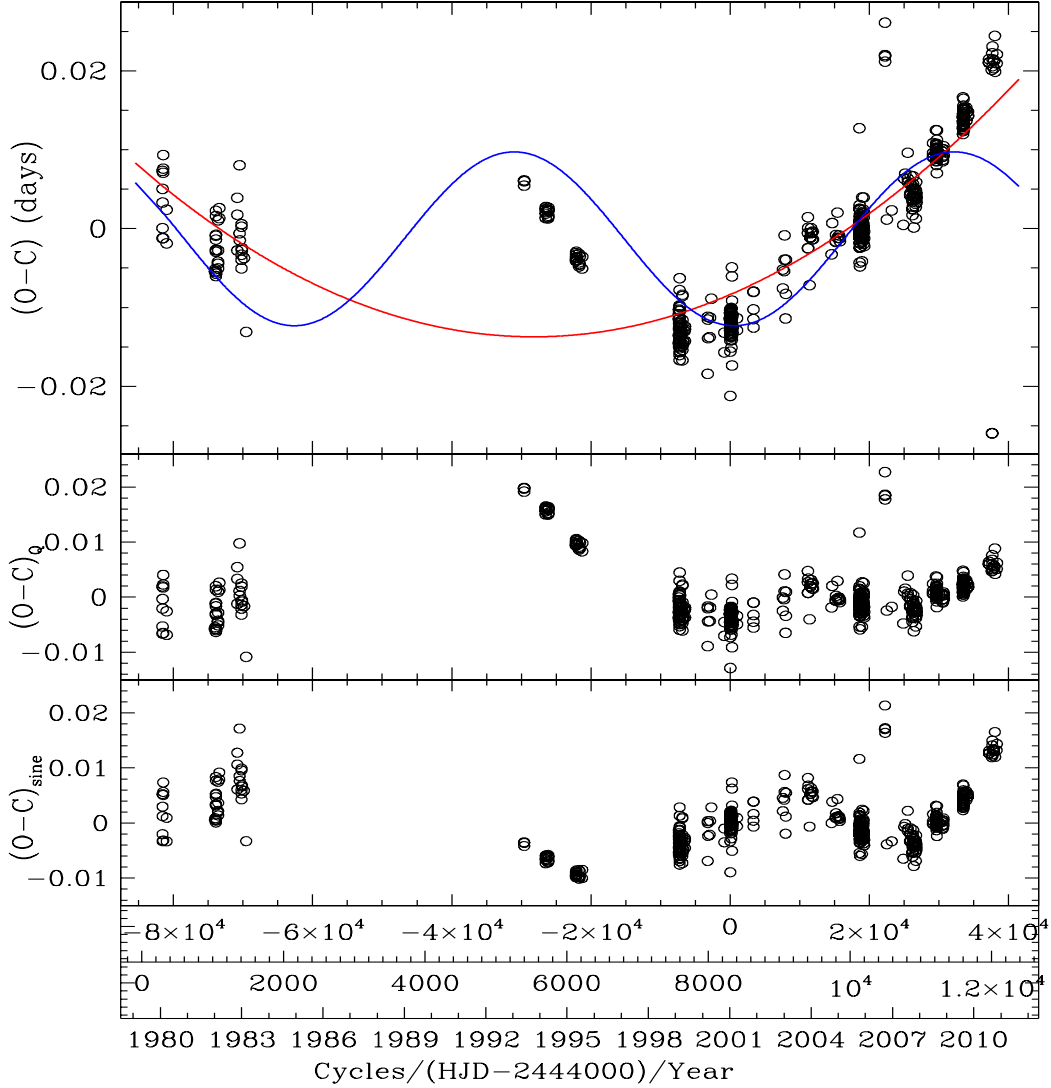}
  \caption{$(O-C)$ diagram of AN Lyn with 541 maxima.
  Top: the $(O-C)$ calculated by the newly derived ephemeris (eq.\ref{eq:new-epoch}) and the fits
of parabola (eq.~\ref{eq:Qfit}) and sinusoid (eq.\ref{eq:O-C-sin});
  Second panel: $(O-C)_{\rm Q}$, the residuals after the parabola removed;
  Third panel: $(O-C)_{\rm sine}$, the residuals after the sine curve removed;
  Bottom: three timescales in cycles, days and years are given for easy reading.}
\label{Fig:O-C-Qsine}
\end{figure}

\placefigure{Fig:O-C-Qsine}

By using eq.~(\ref{eq:new-epoch}),
we calculated the final cycles and corresponding $(O-C)$ residuals for each maximum.
The resultant $(O-C)$ are drawn in Fig.~\ref{Fig:O-C-Qsine},
where the fitting parabola is formulated in the second-order polynomial of
$(O-C)$ on $E$ as
\begin{equation}
  \begin{array}{lrclll}
  (O-C) = &  -0.008361 & + & 3.7945\times 10^{-7} E & + & 6.72\times 10^{-12}  E^{2}\\
          &\pm0.000299 &\pm& 0.1017                 &\pm& 0.20 \\
{\rm i.e.~ }\\
{\rm HJD_{max}}=& 2452307.069058 & + & 0.0982748345\,E & +& 6.72\times 10^{-12}\,E^{2}\\                &
  \end{array}
\label{eq:Qfit}
\end{equation}
with a standard deviation of the fitting residuals of $\sigma$=0\fd00488.
The above quadratic coefficient indicates an increasing period change rate of
$\dot{P}/P= (5.1\pm0.1)\times 10^{-7}$ yr$^{-1}$,
which is consistent with theoretical predictions for evolving main sequence
stars \citep{breg98}.
Thus, the observed long-term period change for AN Lyn may be an evolutionary period change.
If that were so, a normal quadratic fit could account for the distribution of
the $(O-C)$ residuals.
However, as shown, the maxima observed since 1994 are more consistent.
If the fit were just applied to the maxima since 1994, then we would have
$
{\rm HJD_{max}}= 2452307.066801  +  0.0982745949\,E  + 18.8\times 10^{-12}\,E^{2}
$
with a standard deviation of residuals of $\sigma$=0\fd003173.
But this indicates a much higher increasing period change rate of
$\dot{P}/P= (14.3\pm0.1)\times 10^{-7}$ yr$^{-1}$,

There are several maxima with large $(O-C)$ scatter.
Let us look at the four maxima observed in 2008 December at NAOC.
After checking the observing log book, we believe that
they probably suffered from bad timing.
The computer clock of the BFOSC system was not ensured
as the night assistant observers used to pay little attention to
time-series photometry and neither GPS timing nor time synchronization
over the Internet was employed.
Fitting without these four maxima led to
$  (O-C) = -0.008335\pm0.000284 +
3.5826\pm0.1013\times 10^{-7} E + 6.413\pm0.194\times 10^{-12} E^{2}$
with $\sigma$=0\fd004571.
Therefore, these four outlying maxima have little affect
on fitting results, there is no need to exclude them and
thus each maximum datum was equally assigned a full weight in all fittings.
The SNO data in 1994--1996 are obviously deviated from the parabolic fit.

We have noted the most recent $(O-C)$ diagram given by \cite{wils10} who
used a different ephemeris
HJD$_{\rm max}$=2451583.0767+0.098274972$\times E$.
The key issue is that they assumed that the cycles calculated in \cite{hintz05}
may had been miscounted due to the large gap from lack of observations between 1980s and 1990s.
Therefore, they added one cycle to the maxima after 1990.
Similar treatment was also taken by \cite{liq10},
who also believed that the cycles were miscounted. They corrected them by
adding a full period to the $(O-C)$ values of these maxima.
However, the two papers gave inconsistent $(O-C)$ diagrams.
Our consideration of early ephemerides for AN Lyn indicate that these authors' choice of cycle numbers
may be incorrect. The large gap in observations is
a problem, but perhaps does not force upon us those particular choices of cycle count.
We found that taking the ephemeris derived even earlier by \cite{costa84},
one would have an $(O-C)$ graph similar to Fig.~\ref{Fig:O-C-Qsine}.

Considering the light travel effect in a binary orbital motion as a cause of period change,
a sinusoid fit was found directly (see Fig.~\ref{Fig:O-C-Qsine})
\begin{equation}
   (O-C) = -0.0012917  + 0.01102 \sin [2\pi(1.5816\times 10^{-5}\,E +0.74073)]  ~.\\
\label{eq:O-C-sin}
\end{equation}
The analytic standard errors for the three coefficients of the sine term above
are 0.00029, 0.12E$-$5, and 0.00414, respectively.
The standard deviation of residuals with this fit is $\sigma$=0\fd004683.
The frequency value 1.5816E$-$5 equals to a period of about 17 years.
It is worth to note that the recent maxima of high quality
did not turn down to follow the sine curve in Fig.~\ref{Fig:O-C-Qsine}.

\begin{figure}[t]
\vspace{-5mm}
\centering
\includegraphics[width=0.85\textwidth,height=138mm,angle=-0,clip=true]{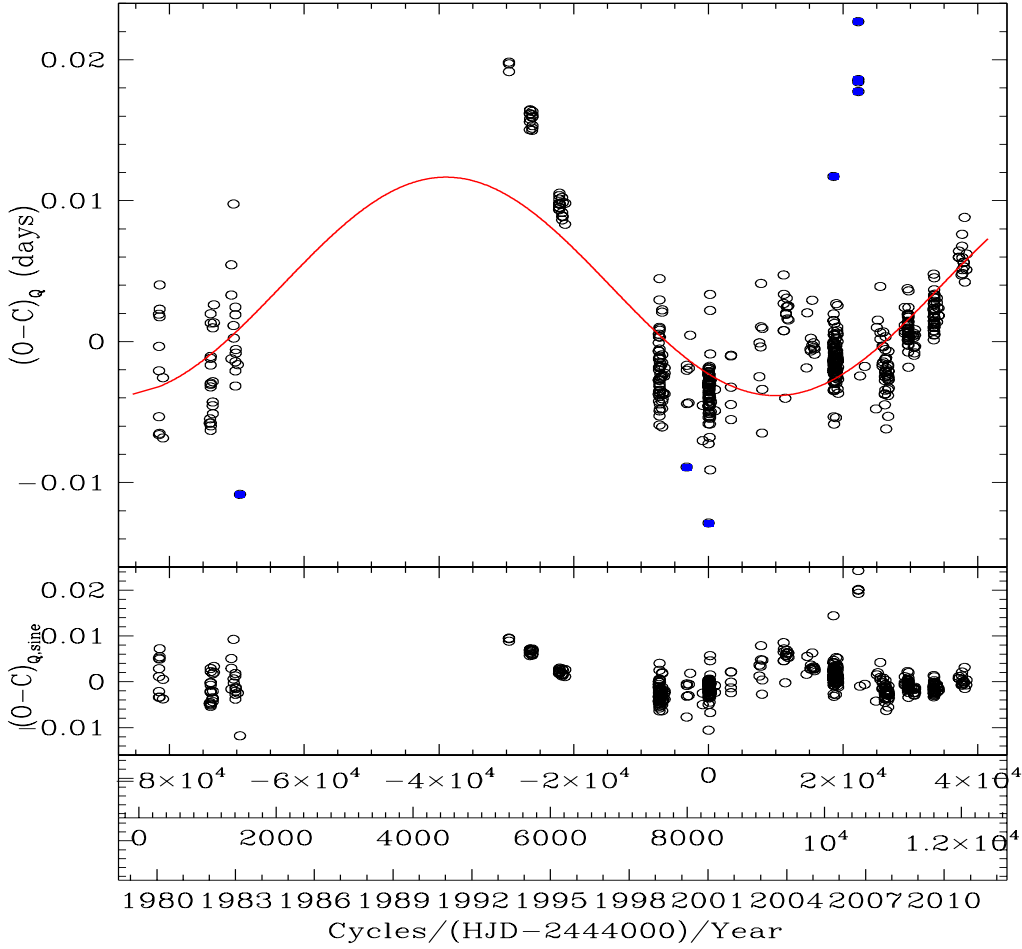}
  \caption{$(O-C)$ diagram of AN Lyn with 541 maxima:
   a forced sinusoid fit to $(O-C)_{\rm Q}$ residuals.
   Several outliers are drawn in blue solid dotss. }
\label{Fig:O-C-sinQ}
\end{figure}

However, if one applied a Fourier transform and a nonlinear least-squares sine fit
to the residuals using the parabolic ephemeris equation~(\ref{eq:Qfit}),
i.e. to the $(O-C)_{\rm Q}$ residuals in Fig.~\ref{Fig:O-C-Qsine}, we would have
\begin{equation}
  {(O-C)_{\rm Q}} = 0.0040125 + 0.00795 \sin [2\pi(1.0359\times 10^{-5}\,E +0.64976)]\\
\label{eq:O-C-sinQ}
\end{equation}
with a standard deviation of residuals of $\sigma$=0\fd00355 (see Fig.~\ref{Fig:O-C-sinQ}).
The frequency value 1.0359E-5 equals to a period of about 26 years.
Now, the recent maxima are well fitted by the sine curve,
which would predict that the maxima in near future go up and then turn down.
We are expecting to see whether the trend is correct or not.

\placefigure{Fig:O-C-sinQ}

\begin{figure}[t]
\vspace{-5mm}
\centering
\includegraphics[width=0.80\textwidth,height=120mm,angle=-0,clip=true]{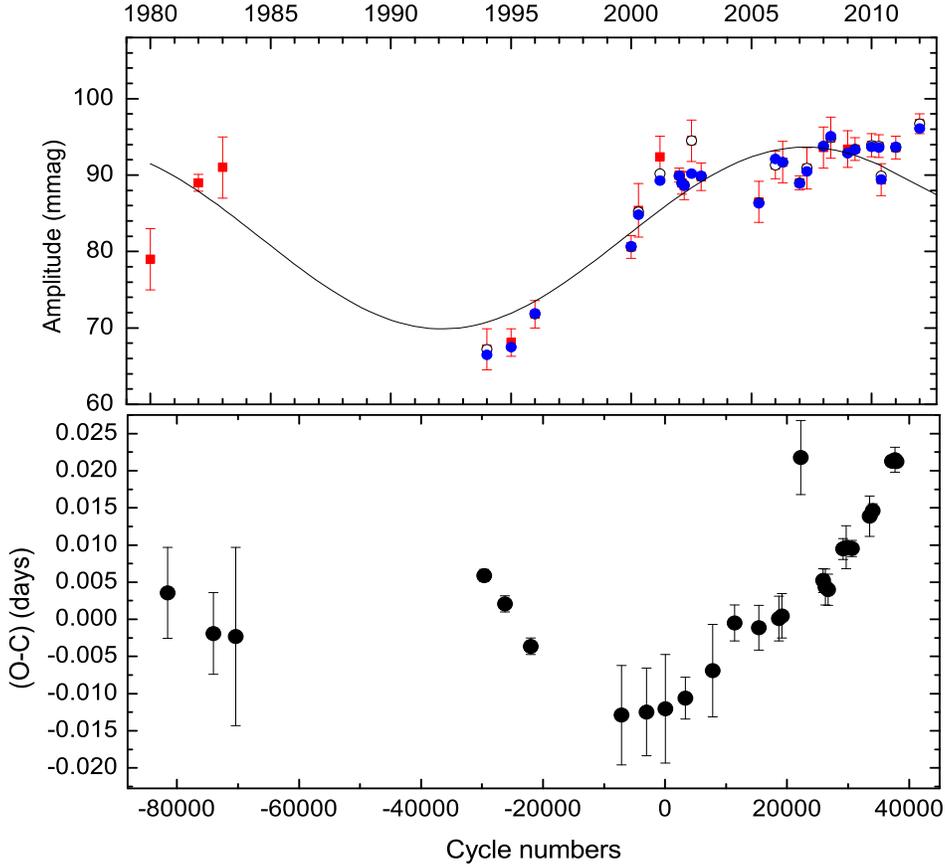}
\vspace{-3mm}
\caption{Binned $(O-C)$ diagram compared with the pulsation amplitudes. }
\label{Fig:omc-ampl}
\end{figure}

In order to check whether the $(O-C)$ pattern matches with the pulsation amplitude variations,
we further determined the medians and error bars for each data bin
in the $(O-C)$ diagram and redrew them along with the amplitudes
in Fig.~\ref{Fig:omc-ampl} for a phase comparison.
They did not run in the same phase.
But their periods assuming a binary orbit are quite close:
30-yr cyclic amplitude variation and 26-yr period change.
To some extent, the sine-wave fits could have over-weighted
the earlier data in 1980s and 1994--1996 because we used equal weights for all points,
but these data suffered from big gaps compared to those since year 2000.
\placefigure{Fig:omc-ampl}

\subsection{Detailed time evolution behaviour}
So far we have checked the averaged long-term pulsation behavior.
Here we try to gain an insight into the temporal evolution of pulsation behavior
by means of the wavelet method.
Wavelets were designed to study how a signal changes over time.
Unlike Fourier analysis, wavelet analysis can give the instant frequency of a signal
at a localized time.
So wavelet analysis is an especially useful technique for
looking at stars whose periods, amplitudes, and/or modes change over measurable timescales.
Our extensive data of AN Lyn provide an excellent opportunity to examine
the time-dependent behaviour of its pulsation,
e.g. the phenomenon of amplitude variability and the changes in period.
As we did in previous works \citep[e.g.][]{zhoq03},
the weighted wavelet Z-transform
(WWZ\footnote{The continuous wavelet transform of a function of time $x(t)$ is defined as:
$W(\omega,\tau , x(t)) = \omega^{0.5}\int x(t) f^{*} (\omega,(t-\tau )) {\rm d}t$
\citep[eq.1-1,1-4][]{fost96}.
The WWZ algorithm performs the wavelet
transform using an analyzing wavelet function
$f(\omega,(t-\tau )) = e^{i\omega(t-\tau)-c \omega^2(t-\tau)^2}$,
which includes both a periodic, sinusoidal test function $e^{i\omega (t-\tau )}$
and a Gaussian window function $e^{-c \omega^2(t-\tau)^2}$,
where both the frequency $\omega$ and the user defined constant c determine
the width of the window. The algorithm fits a sinusoidal wavelet to the data,
but as it does so, it weights the data points by
applying the sliding window function to the data \citep[refers to][]{temp04}.})
developed by \cite{fost96} was used in this wavelet analysis.

\begin{figure}[thp!]
\vspace{0mm}
\centering
\includegraphics[width=0.8\textwidth,height=75mm,angle=-0,clip=true]{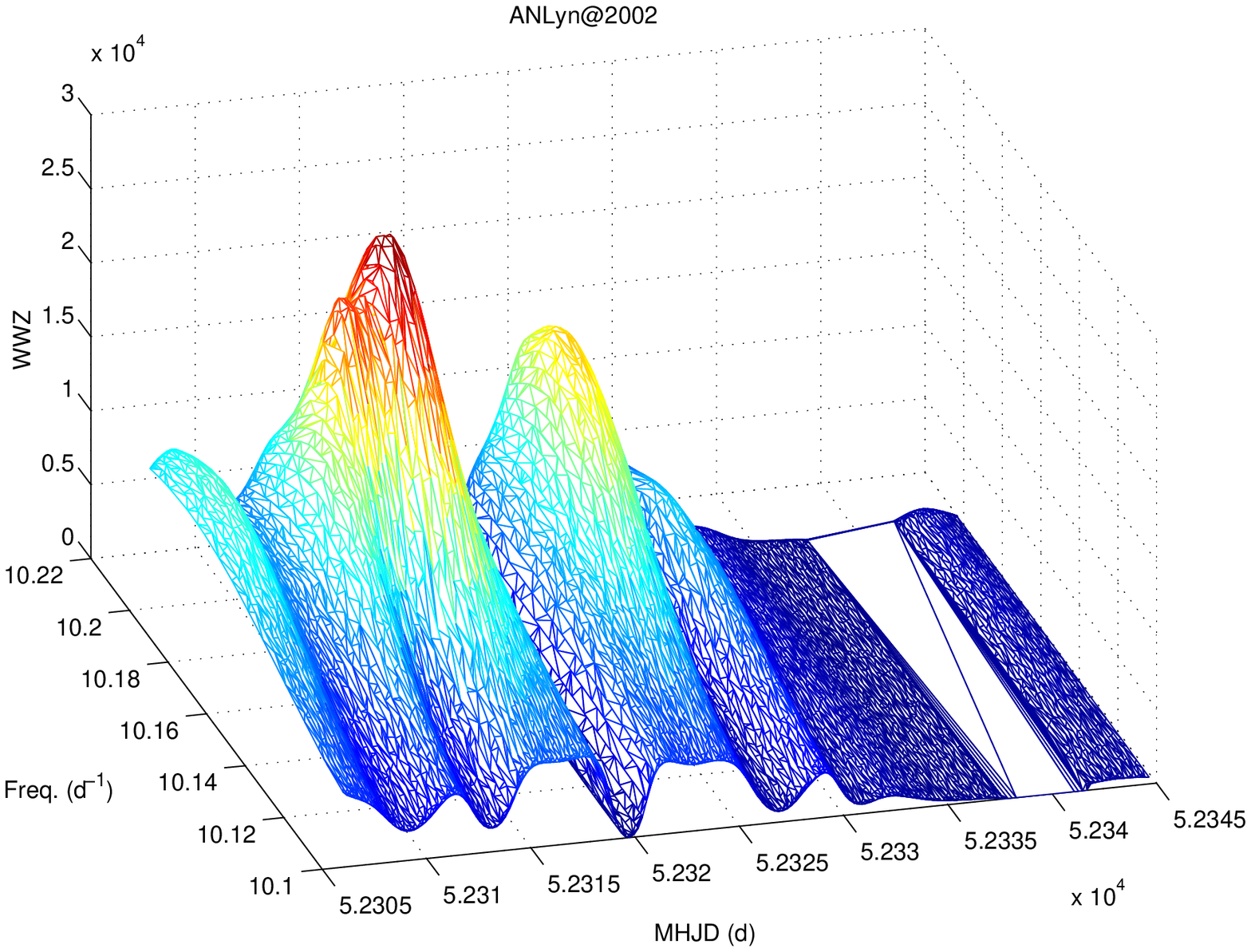}
\includegraphics[width=0.8\textwidth,height=75mm,angle=-0,clip=true]{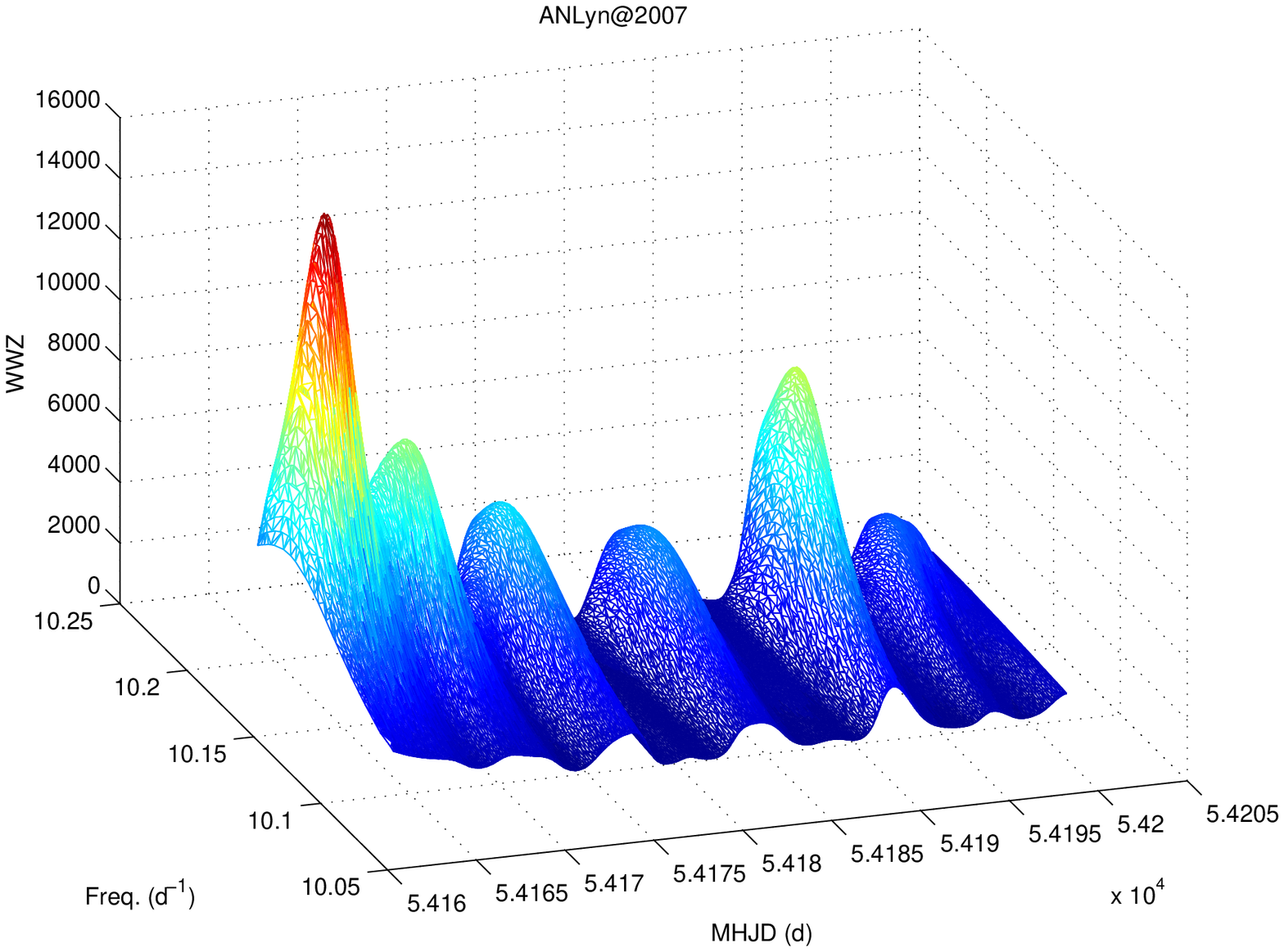}
\includegraphics[width=0.8\textwidth,height=75mm,angle=-0,clip=true]{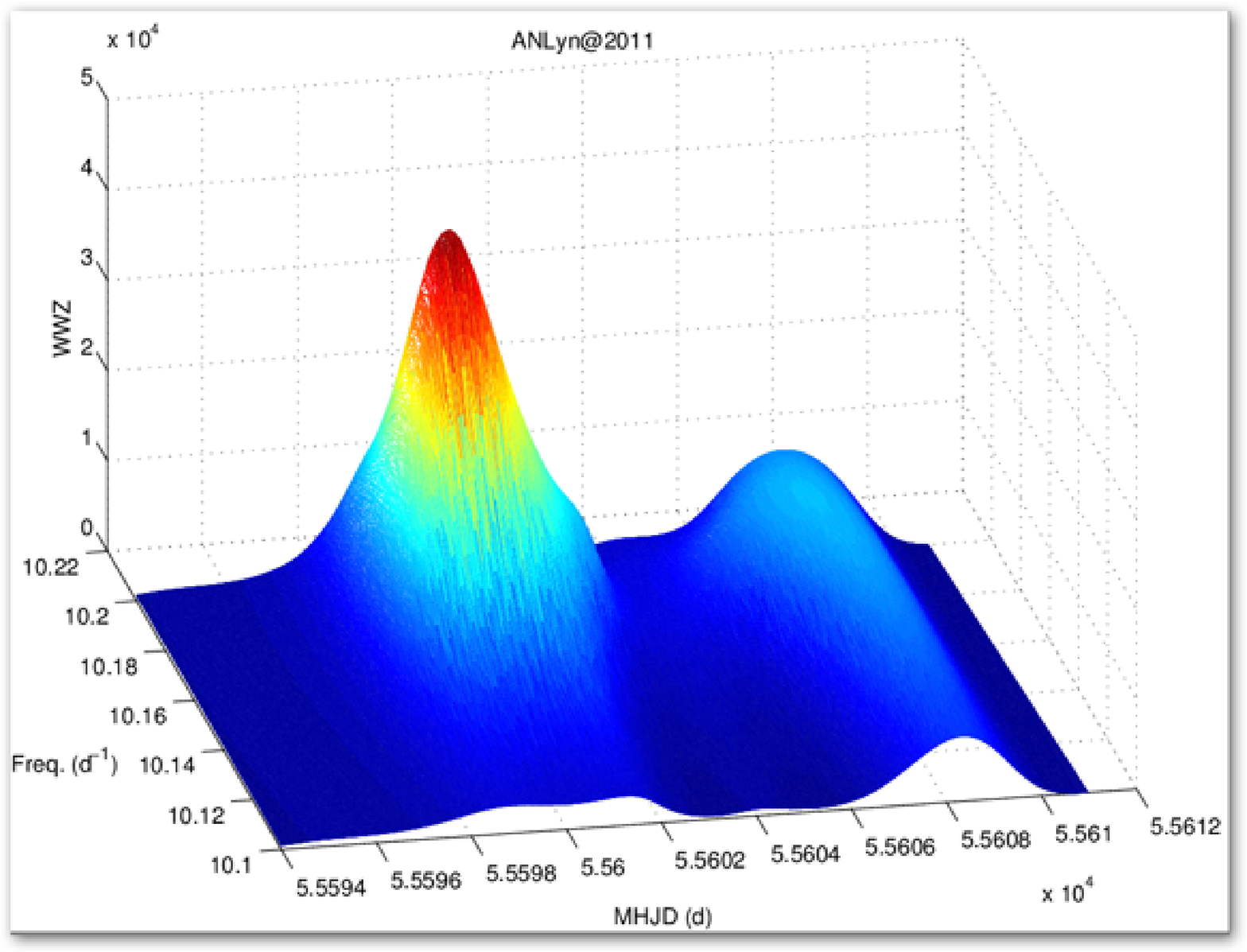}
\caption{Time-frequency map of AN Lyn's pulsation --- the weighted wavelet Z-transform,
in three representative observing seasons of 2002, 2007 and 2011.}
\label{Fig:wavelet-3D}
\end{figure}

However, the WWZ program might not yield what one expects
if unsuitable control parameters were used.
By trial and error, we finally found the decay constant of 0.000125,
which gives a wider sliding window, better period/frequency resolution and lower temporal resolution.
We set frequency in the interval of 10.0 to 10.3\,d$^{-1}$ covering
the main frequency $f_1$=10.1756 d$^{-1}$
and the range of possible changes.
This also means a tiny frequency variation up to 0.2\,d$^{-1}$ was allowed.
The time was set to vary from the beginning to the end of a data set.
Frequencies were calculated in a step of 0.001\,d$^{-1}$.
As the wavelet calculation usually terminated when encountering a big gap,
and the fact that there are many large gaps in the data,
wavelet analyses were separately applied to individual seasonal data sets.
On the other hand,
in the unobserved gaps, the amplitude and frequency show large fluctuations.
Therefore, we intend to first analyse the consecutive data sets without big gaps.
In Fig.~\ref{Fig:wavelet-3D}, we present
the whole pulsational time evolution for
the three representative observing seasons 2002NAOC, 2007NAOC and 2011 ($V$ of $BVI$)
as function of both time and frequency
in a manner of three-dimension mesh graphs.

Now by decomposing the time series light curves into time-frequency space,
we are able to reveal which pulsation frequencies are dominant in certain time intervals.
For example,
at HJD 2454166.0312, $f$= 10.176204\,d$^{-1}$ with amplitude of 0\fm08959 and WWZ value of 15700.0 and
at HJD 2454192.7422, $f$= 10.177305\,d$^{-1}$ with amplitude of 0\fm09208 and WWZ value of 8740.6 (the two highest peaks in 2007NAOC group), and
at HJD 2455599.8203, $f$= 10.175031\,d$^{-1}$ with amplitude of 0\fm09764 and WWZ value of 45616.9 (the highest peak in 2011 data set).
The visual representation of the wavelet analysis in Fig.~\ref{Fig:wavelet-3D},
depicts that
the peaks of WWZ amplitudes correspond roughly to the occurrence of high-amplitude cycles.
The periodicity is evident --- the individual peaks indicating instantaneous dominant pulsation
are not at the same frequency at different instants, that is the period is not constant.
Using this visualization, changes in frequency and amplitude can be observed easily.
We got highly variable results within the time interval of a data set.
However, within the time intervals involved in the demonstrated data sets,
no regularity of variation can be derived.

\begin{figure}[thp!]
\vspace{0mm}
\centering
\includegraphics[width=0.8\textwidth,angle=-0,clip=true]{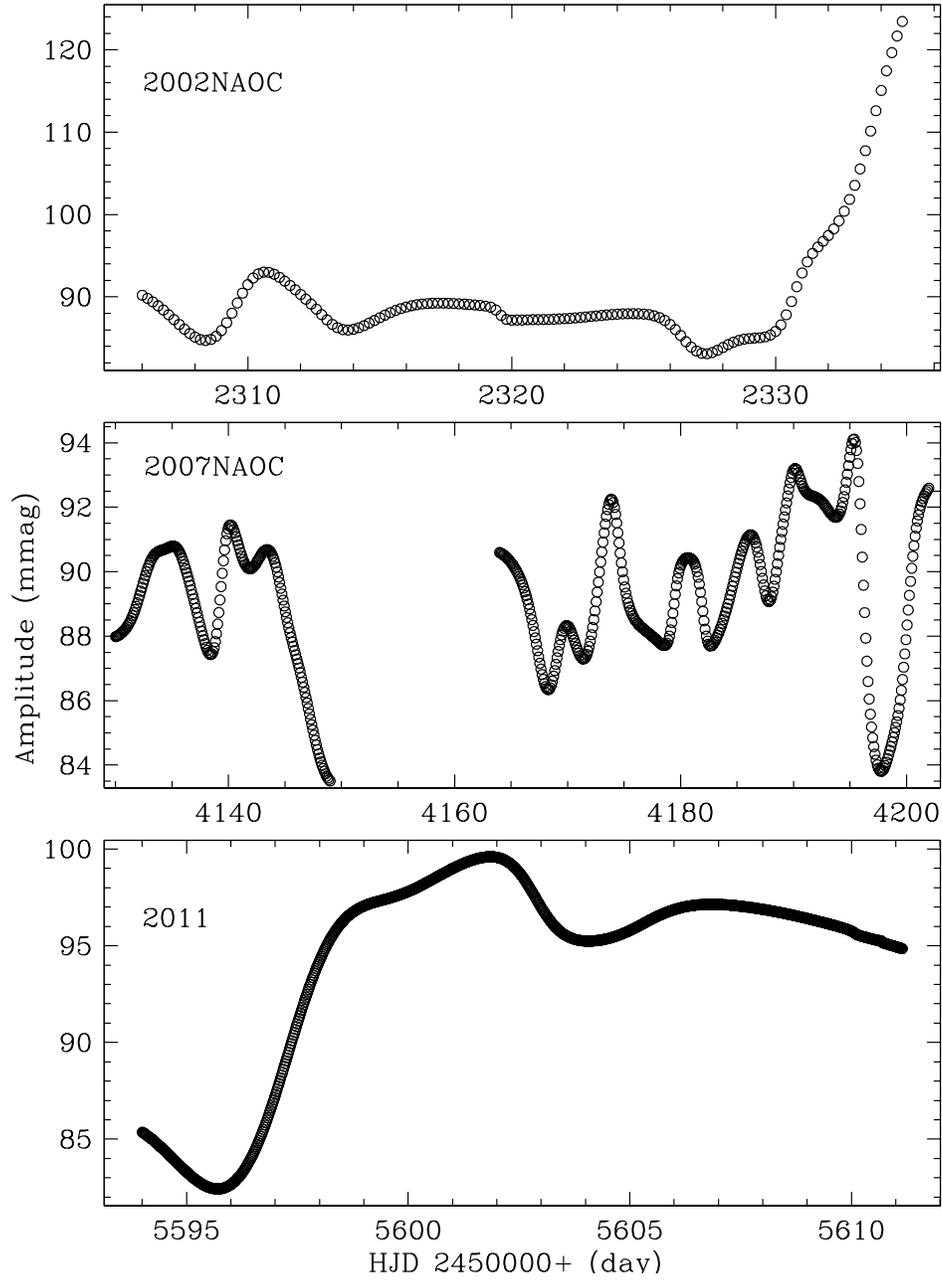}
\caption{Pulsation amplitudes evolution of AN Lyn in the three representative observing seasons.}
\label{Fig:wavelet-ampl}
\end{figure}

\begin{figure}[thp]
\vspace{0mm}
\centering
\includegraphics[width=0.8\textwidth,angle=-90,clip=true]{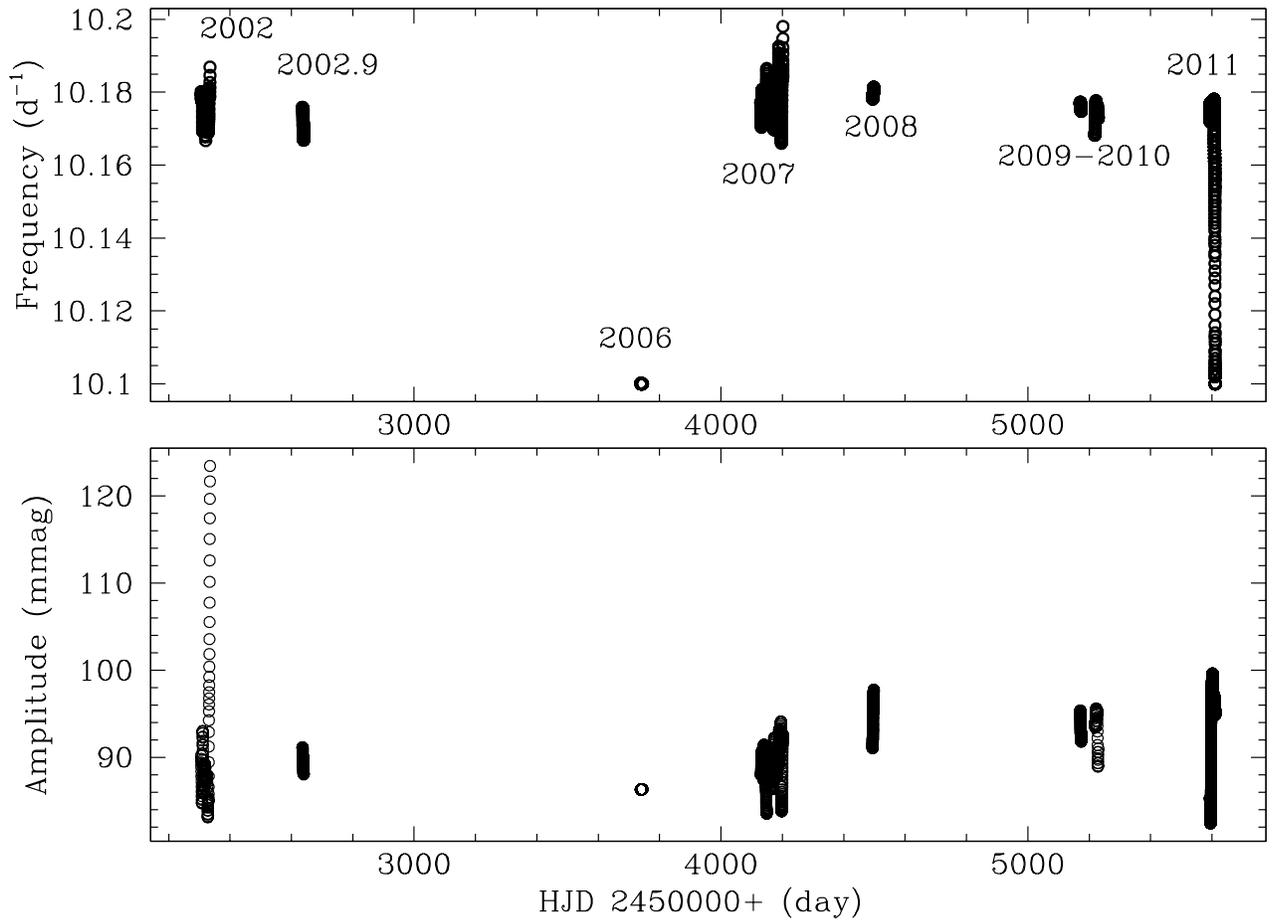}
\caption{Amplitudes and frequencies of AN Lyn evolve with time over 2002--2011 years.}
\label{Fig:wavelet-fa}
\end{figure}

In Fig.~\ref{Fig:wavelet-ampl}, we show the the pulsation amplitude evolution
for these three seasons.
The pulsation amplitudes were derived from the WWZ values.
The ending parts in 2002NAOC and 2007NAOC data sets indicate anomalous cycles.
After checking the corresponding light curves, we make sure it is true:
light curves in HJD 2452334.1--34.4 are anomalous --- should be resulted from bad weather.
The unobserved gap between HJD 2454148.336 and 54164.962 in 2007NAOC group was not calculated.
The wavelet calculation was terminated at HJD 2455607.6 until the end (2455666.496)
of data set 2011 due to a gap of 43 days,
so no wavelet results for the observations in 2455650.4--2455666.496 interval.
To put the wavelet analysis results of individual subgroups together,
we show the instant pulsation amplitude and frequency
as function of time for the main observing seasons in Fig.~\ref{Fig:wavelet-fa}.
The frequencies were determined from the peaks' position of WWZ.
The time dependence of the pulsation amplitudes and frequencies of AN Lyn is clearly resolved.

\clearpage
\placefigure{Fig:wavelet-3D}      
%

\placefigure{Fig:wavelet-ampl}      
%

\placefigure{Fig:wavelet-fa}      

\clearpage
\section{Ending remarks}
\label{sect:Conclusion}

Aiming towards a time-evolution study of pulsation of interesting stars,
we have investigated the temporal-dependent behaviour of the pulsation
of the $\delta$ Scuti star AN Lyncis
based on the data collected through 2001 to 2012.
The data were mainly obtained from twice tri-continent campaigns in 2002 and 2011.
Making use of the new data and those in the literature,
the period and amplitude variability have been examined
on an extended time base over twenty years.
Present analyses show that AN Lyn is undertaking long-term cyclic variations in
both amplitude and period over the time in question.
\cite{zhou02} stated that
``if the cyclic amplitude is real, a similar cyclic variation would be present
in the $(O-C)$ diagram with the same timescales as for the amplitude modulation".
Our new results show that the pulsation amplitude of AN Lyn more or less varies
in a sinusoid wave of period about 30 years, and that
the primary pulsation period changes as the star moves in a binary orbit of period about 26 years.
That is the period change due to light-time effect in a binary system.
Figure~\ref{Fig:ampl} exposes the deviation of the most recent amplitude in 2012 from the sine curve.
The amplitudes in 2012 were measured individually for three subsets of data,
but they all are the same of about 96.5 mmag and higher than any other seasonal values.
Clearly, continued observations are needed to examine whether or not
the pulsation amplitude turns onto a new increasing path.

Sinusoid fittings to both $(O-C)$ and amplitude patterns may be overwhelming
in explaining the cause of variations.
The most recent observations did not follow the sinusoidal fitting trend (see
Figs.~\ref{Fig:ampl} and ~\ref{Fig:O-C-Qsine}).
As \citet{hintz10} found that
AN Lyn was not turning over or down at the previously reported point of a sine curve.
Nevertheless, the sinusoid-like pattern is a better fit to the $(O-C)$ residuals when
a parabola representing evolutionary effect in the period change was removed.
Figure~\ref{Fig:O-C-sinQ} shows the $(O-C)_{\rm Q}$ residuals and the sinusoid fit.
As such, the orbital motion of AN Lyn being a binary system might have induced
both the pulsation amplitude and period variation in the manner shown in Fig~\ref{Fig:omc-ampl}.
Consequently, AN Lyn being a binary system cannot be disproved.

On the other hand,
disagreement between the data and fits cannot be neglected concerning the $(O-C)$ and
amplitude variability patterns.
So we believe that orbital motion is still not an exclusive explanation for
both the $(O-C)$ morphology and amplitude variation pattern.

Amplitude variability in $\delta$ Sct stars is common,
we have noted a similar sinusoidal amplitude variation appearing in
the $\delta$ Sct star V1162 Ori \citep{are01}.
Although it is not pure sinusoid,
AN Lyn is another example exhibiting high degree amplitude variability.

A statistics on all the data given in Fig.~\ref{Fig:2002-2012LC-stat} shows that most data points are well within
a range from $-0\fm1$ to $0\fm1$, a few beyond this range could have caused unusual
or anomalous cycles.
A further autocorrelation of 2007--2011 data indicates
how much the star's light variations at late time $(t+\tau)$ differ from that at time {\em t}.
Actually, the pulsation was not simply repeated but was varying with time
in both amplitude and period (see Fig.~\ref{Fig:autocorr}).

By means of wavelet analysis, we have explored the details of the pulsation evolution.
Figure~\ref{Fig:wavelet-3D}, the time-frequency map of the pulsation,
gives us a three-dimension image of the pulsation
changing as function of both time and period/frequency.
This graph reveals instantaneous pulsation information, that is
the different dominant pulsation frequencies at different times
as well as the occurrence of high-amplitude cycles.
Figures~\ref{Fig:wavelet-ampl} and~\ref{Fig:wavelet-fa} exhibit
the instant pulsation amplitude and frequency as function of time.

As for a static analysis, such as by Fourier transform,
and further use of the color photometric data for possible mode identification,
we plan to report our results in another individual article.
By this paper, we publish all the available light curves data and times of light maximum
in appendix for downloading for interested readers.

\begin{acknowledgements}
This work was supported by the National Natural Science Foundation of China (NSFC) and
the Junta de Andaluc\'{\i}a and the Direccion General
de Investigacion (DGI) under project 00-0000. HAS thanks the U.S. National
Science Foundation for support under grant AST-0707756.

\end{acknowledgements}

\appendix
\section{Appendix materials}
Data tables in text and figures in PDF format are available online:\\
1. Table 1: Observing log of photoelectric and CCD photometry of AN Lyn\\
--- LaTeX:~ \url{http://journal-web/ANLyn-table1.tex};\\
--- PDF:~ \url{http://journal-web/AnLyn-table1.pdf}.\\
2. Table 2: List of 306 new times of maximum light of AN Lyn\\
--- LaTeX: \url{http://journal-web/ANLyn-table2.tex};\\
--- PDF:  \url{http://journal-web/ANLyn-table2.pdf}.\\
3. Table 3: List of 541 maxima analyzed in present work\\
--- LaTeX: \url{http://journal-web/ANLyn-table3.tex};\\
--- PDF:  \url{http://journal-web/ANLyn-table3.pdf}.\\
4. Table 4: Pulsation amplitudes of AN Lyn between 1980 and 2012\\
--- LaTeX: \url{http://journal-web/ANLyn-table4.tex};\\
--- PDF:  \url{http://journal-web/ANLyn-table4.pdf}\\
5. Figure 14--23: Graphs for all the new light curves of AN Lyn:\\
--- PDF: \url{http://journal-web/ANLyn-LC.pdf}

\begin{figure}[tp!]
  \vspace{-5mm}
\includegraphics[width=150mm,height=220mm,angle=-0,clip=true]{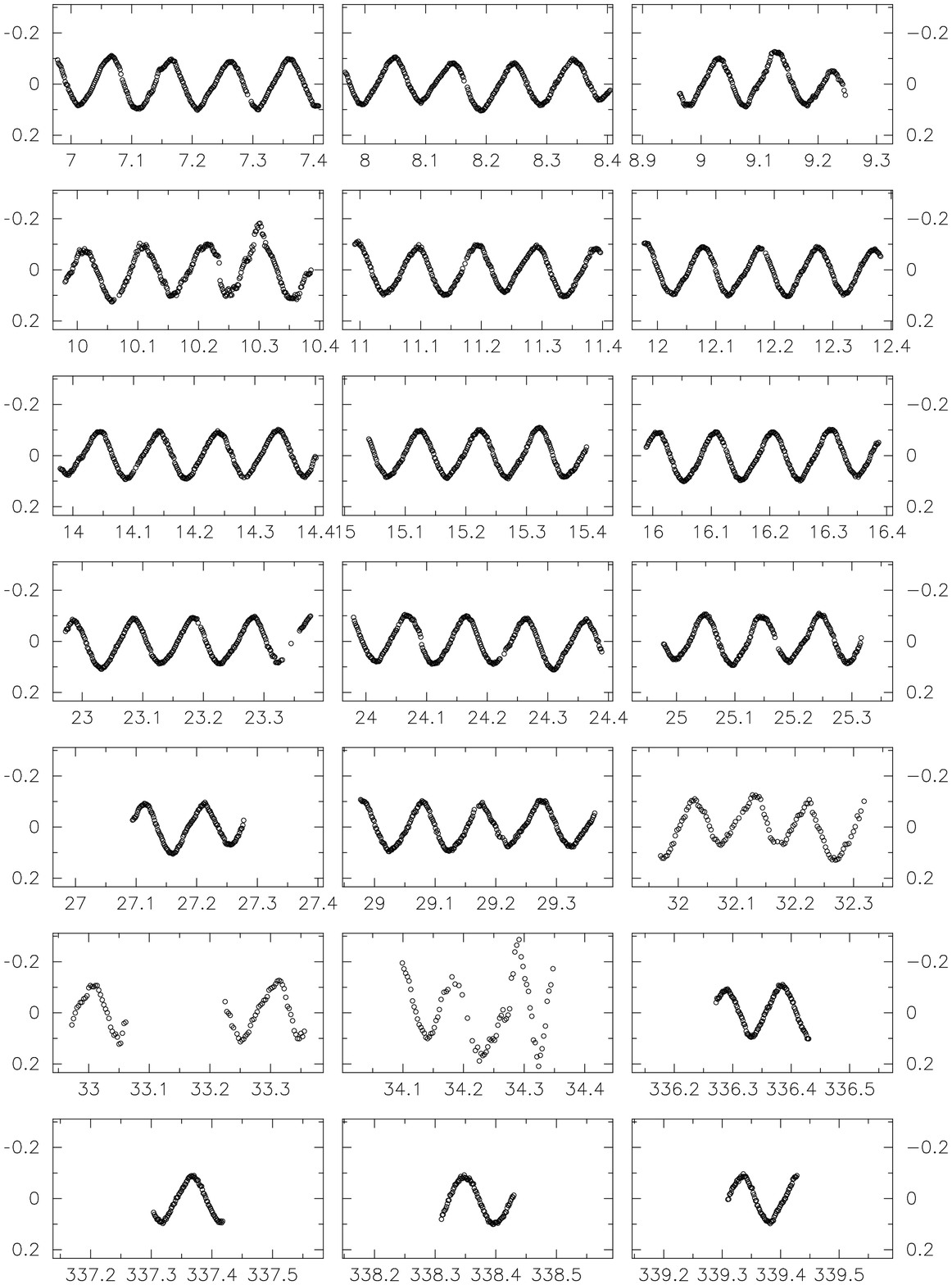}
  \caption{Light curves of AN Lyn: NAOC 2002 in $V$ filter. Integration of 120 seconds
binned from raw 10-s records.  Time in HJD 2452300+ (days). }
\label{Fig:LC2002NAOC}
\end{figure}

\clearpage
\begin{figure}[tp!]
  \vspace{-15mm}
\includegraphics[width=160mm,height=240mm,angle=-0,clip=true]{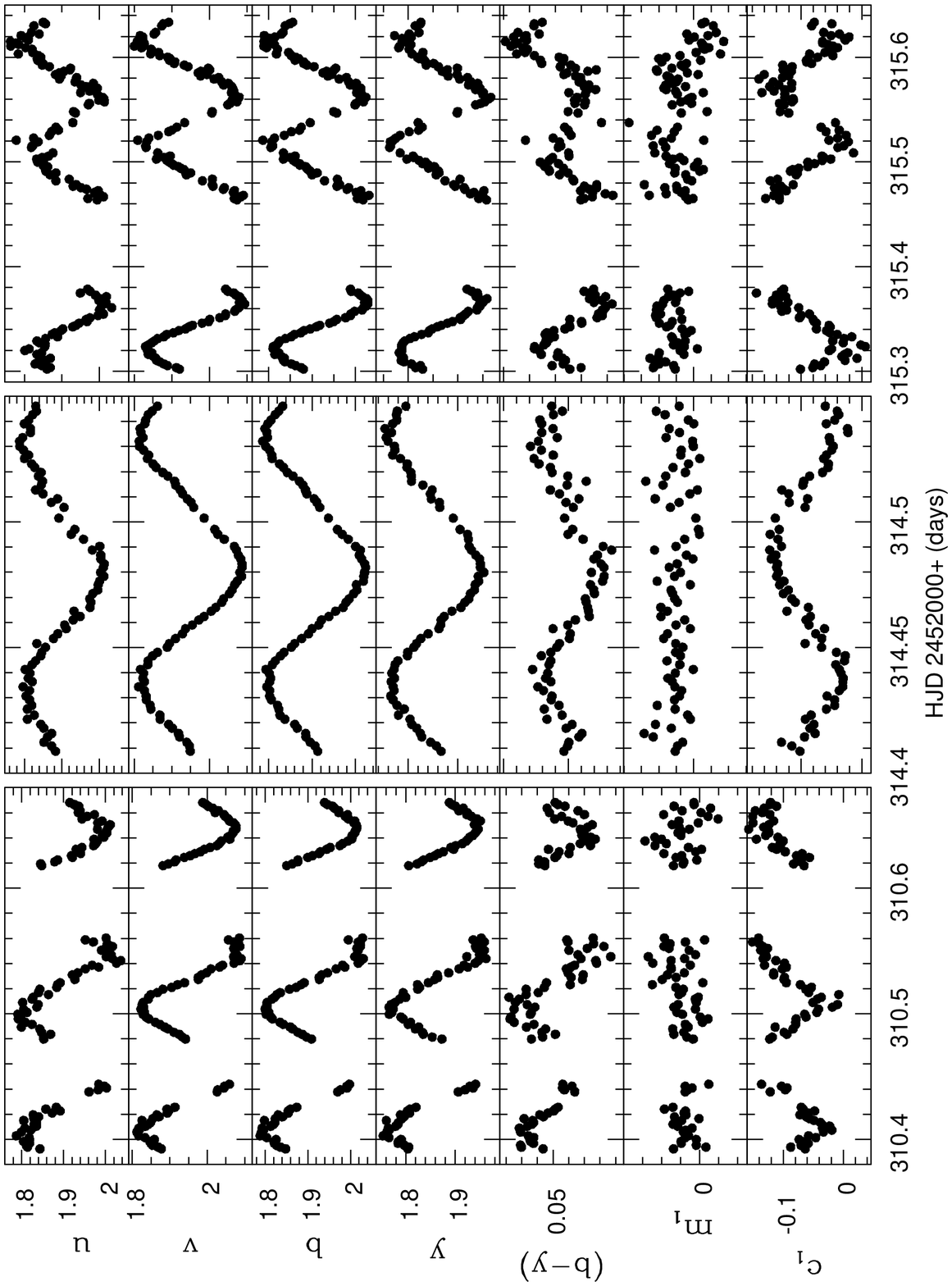}
  \caption{Light curves of AN Lyn: SNO 2002.  }
\label{Fig:LC2002SNO}
\end{figure}
\addtocounter{figure}{-1}
\begin{figure}[tp!]
  \vspace{-15mm}
\includegraphics[width=160mm,height=240mm,angle=-0,clip=true]{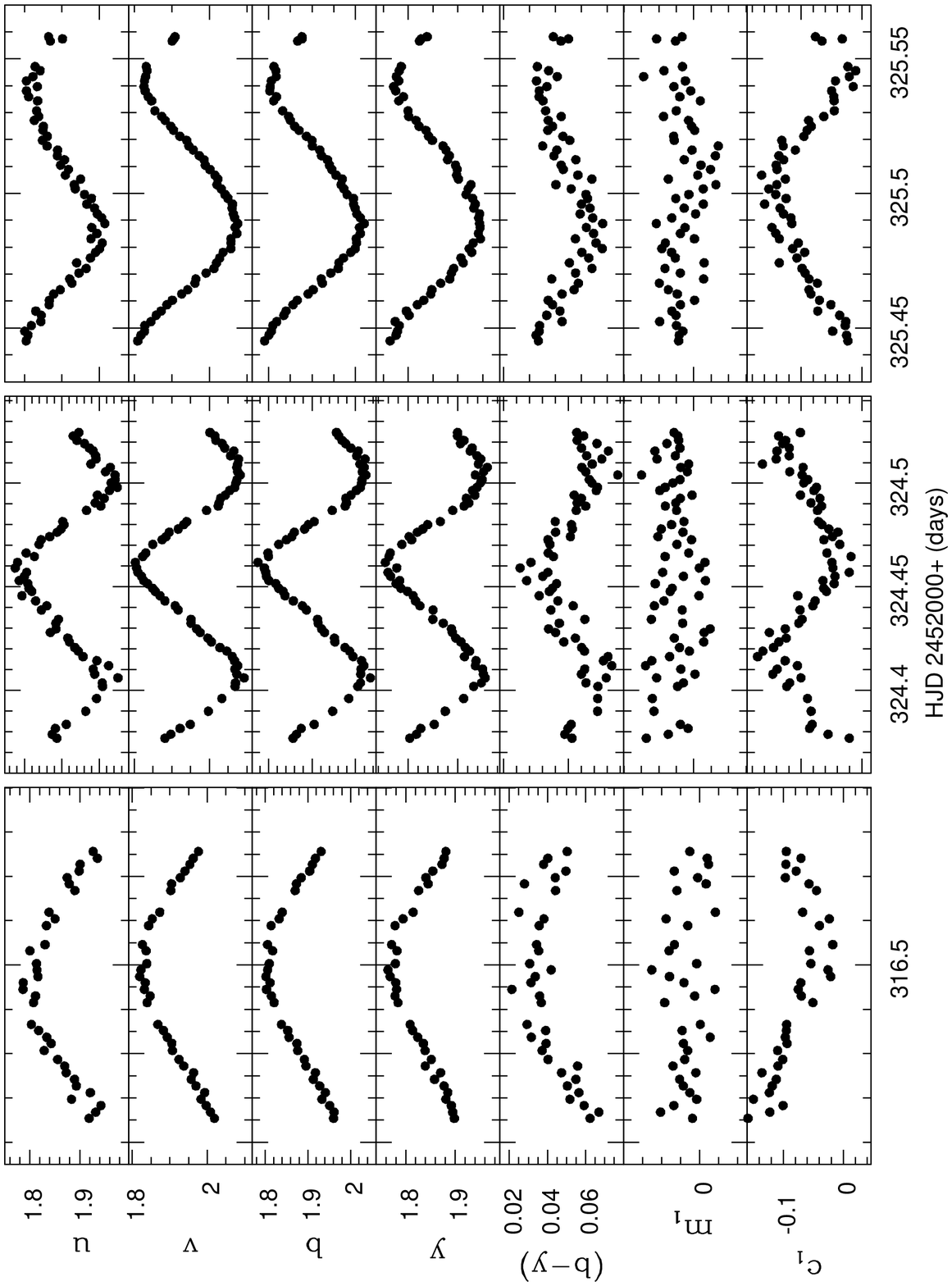}
  \caption{Light curves of AN Lyn: SNO 2002 --- continued.  }
\end{figure}
\addtocounter{figure}{-1}
\begin{figure}[tp!]
  \vspace{-15mm}
\includegraphics[width=160mm,height=240mm,angle=-0,clip=true]{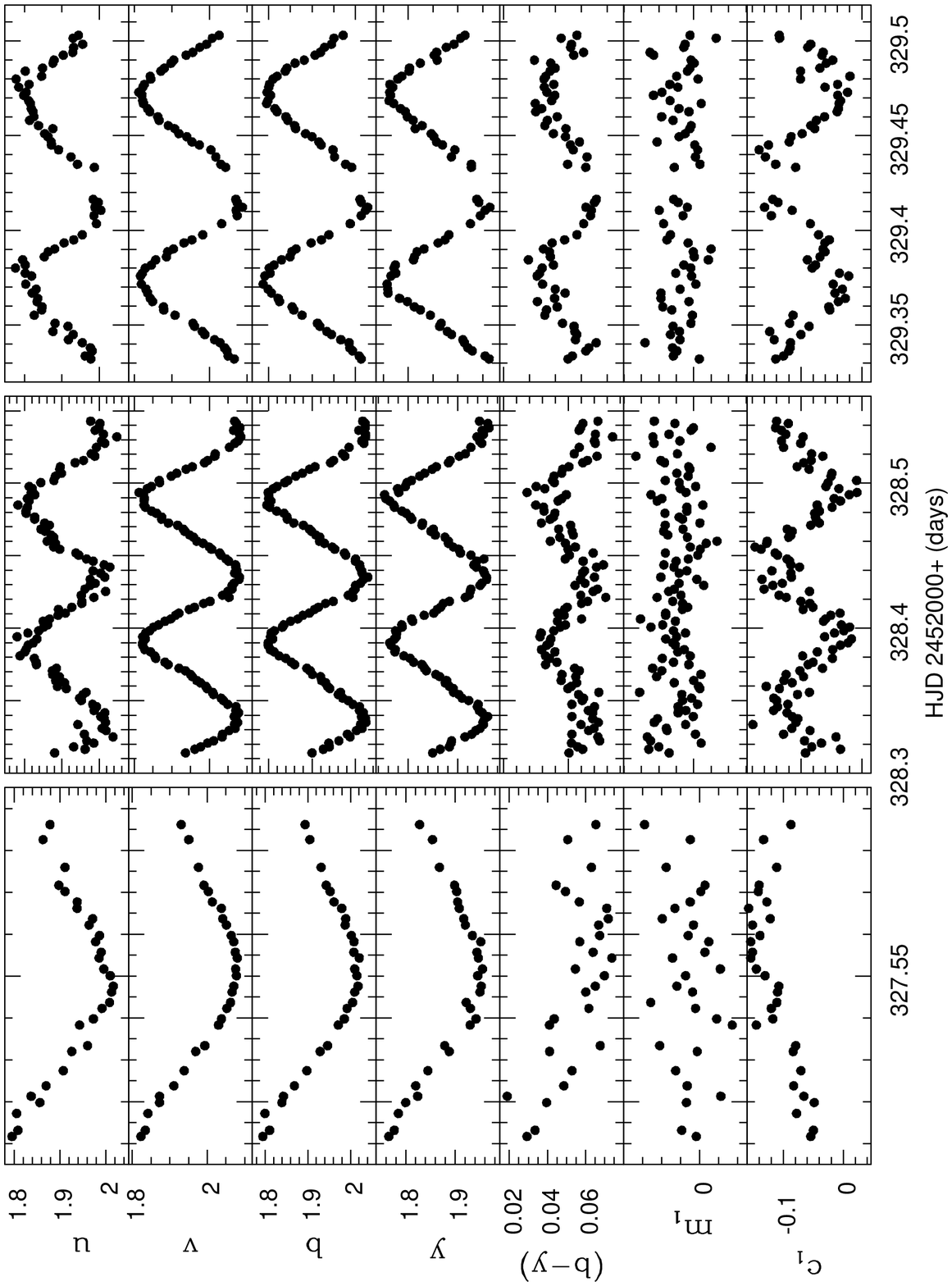}
  \caption{Light curves of AN Lyn: SNO 2002 --- continued.  }
\end{figure}
\addtocounter{figure}{-1}
\begin{figure}[tp!]
  \vspace{-15mm}
\includegraphics[width=160mm,height=240mm,angle=-0,clip=true]{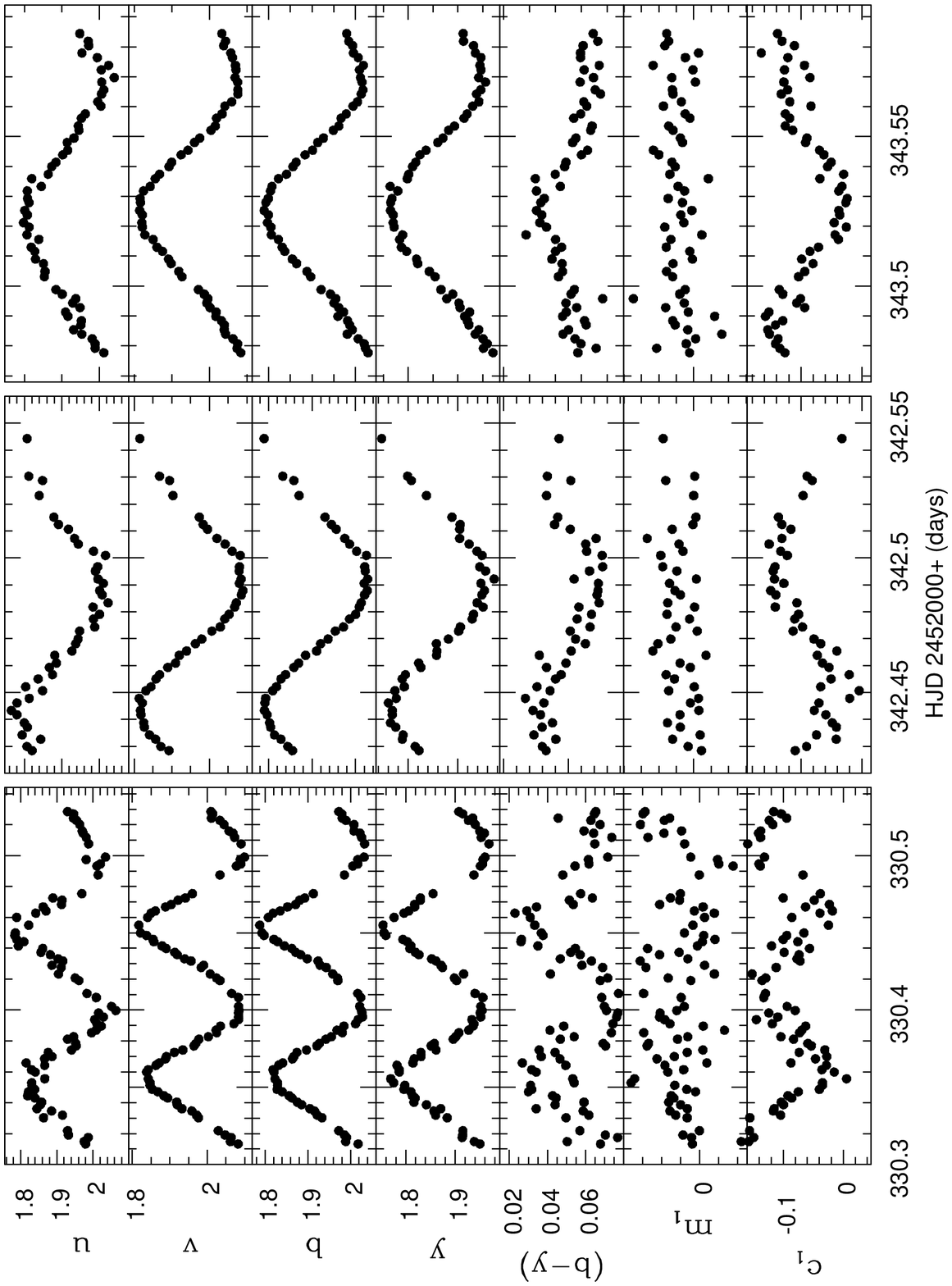}
  \caption{Light curves of AN Lyn: SNO 2002 --- continued.  }
\end{figure}

\clearpage
\begin{figure}[tp!]
  \vspace{-5mm}
\includegraphics[width=150mm,height=220mm,angle=-0,clip=true]{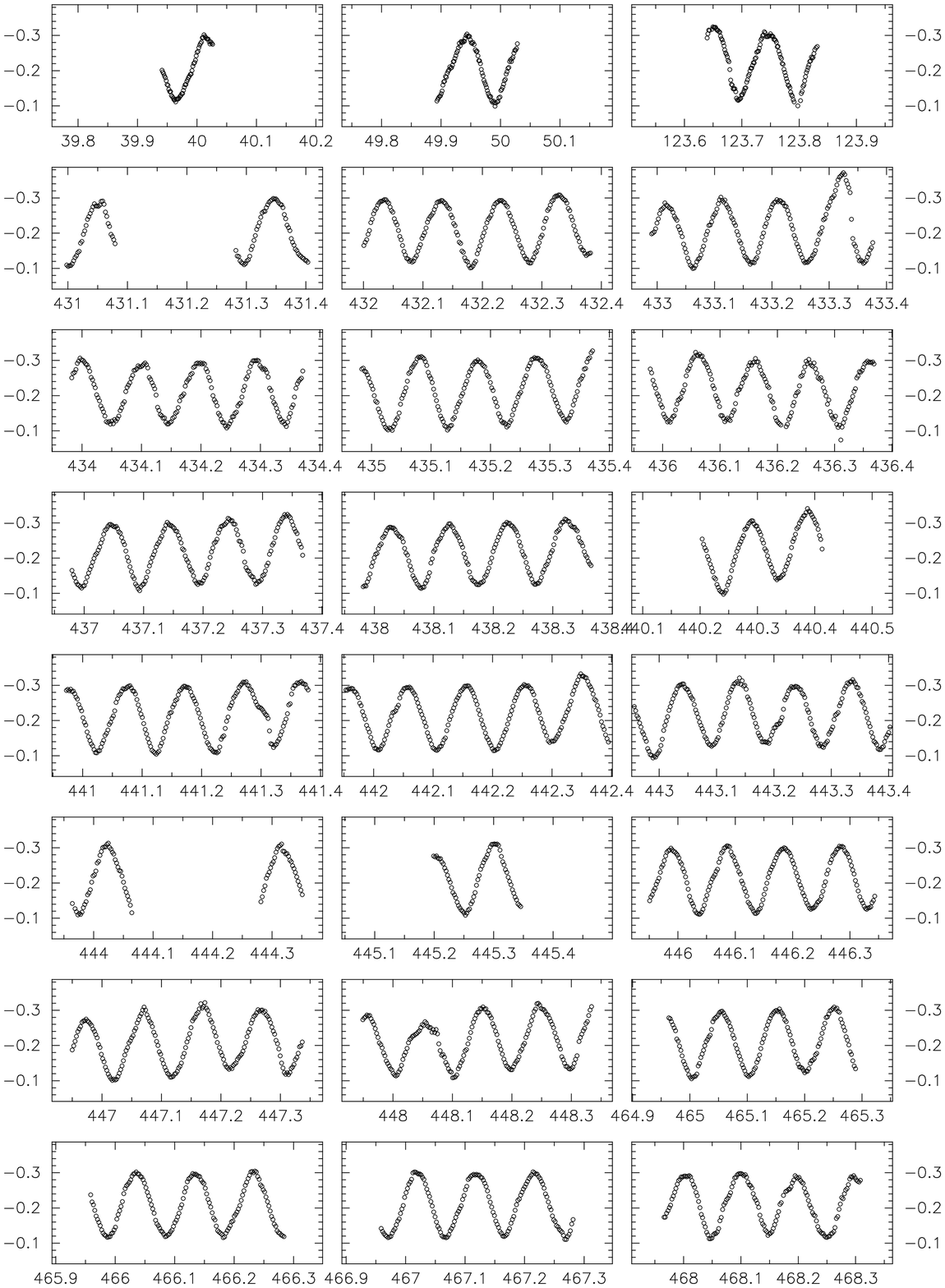}
  \caption{Light curves of AN Lyn: Baker and NAOC 2006--2007 in $V$ filter. Time in HJD 2453700+ (days). }
\label{Fig:LC2007NAOC}
\end{figure}
\addtocounter{figure}{-1}
\begin{figure}[tp!]
  \vspace{-5mm}
\includegraphics[width=150mm,height=220mm,angle=-0,clip=true]{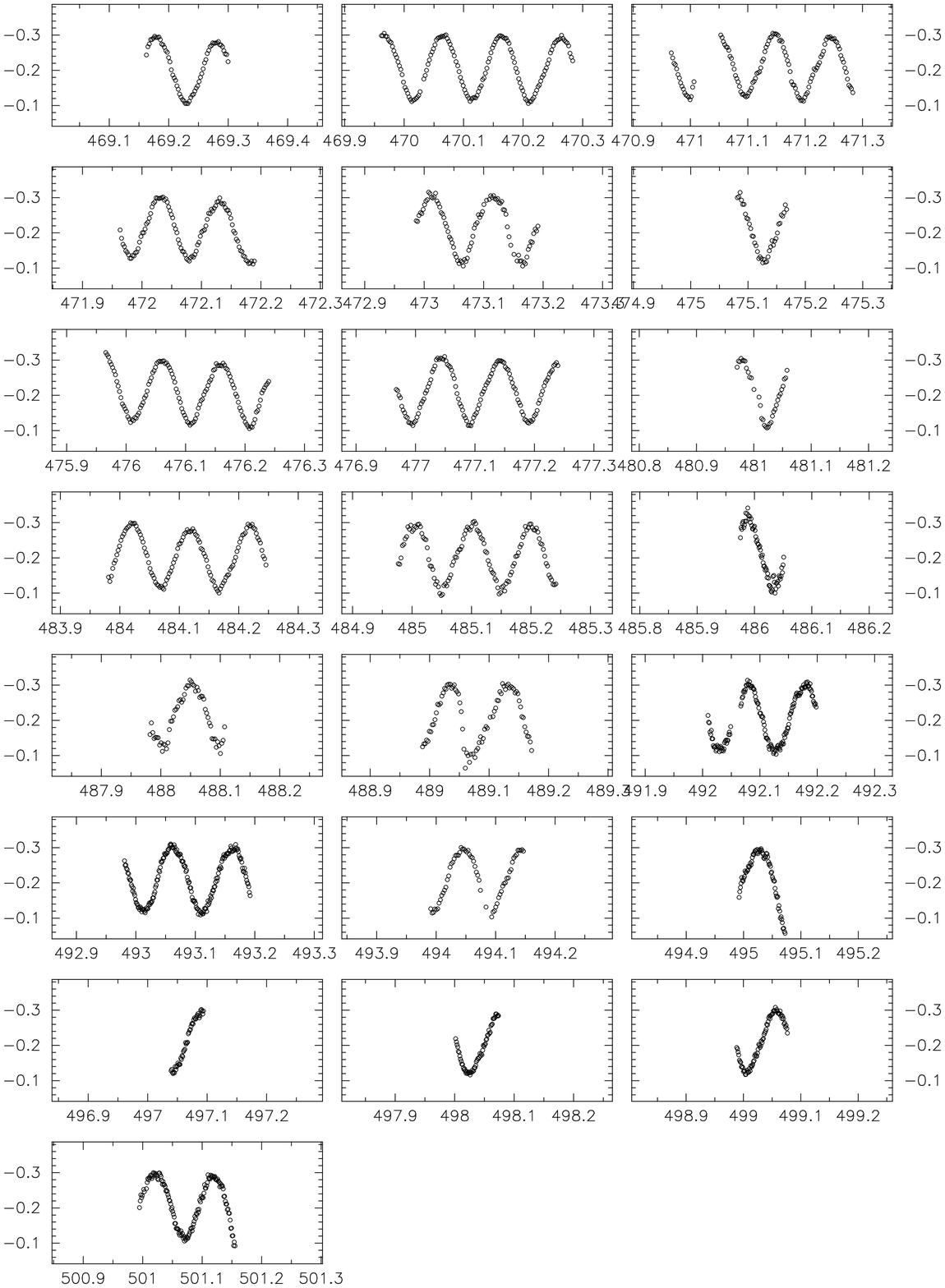}
  \caption{Light curves of AN Lyn: NAOC 2006--2007 in $V$ filter. Time in HJD 2453700+ (days) --- continued. }
\end{figure}

\begin{figure}[tp!]
  \vspace{-5mm}
\includegraphics[width=150mm,height=220mm,angle=-0,clip=true]{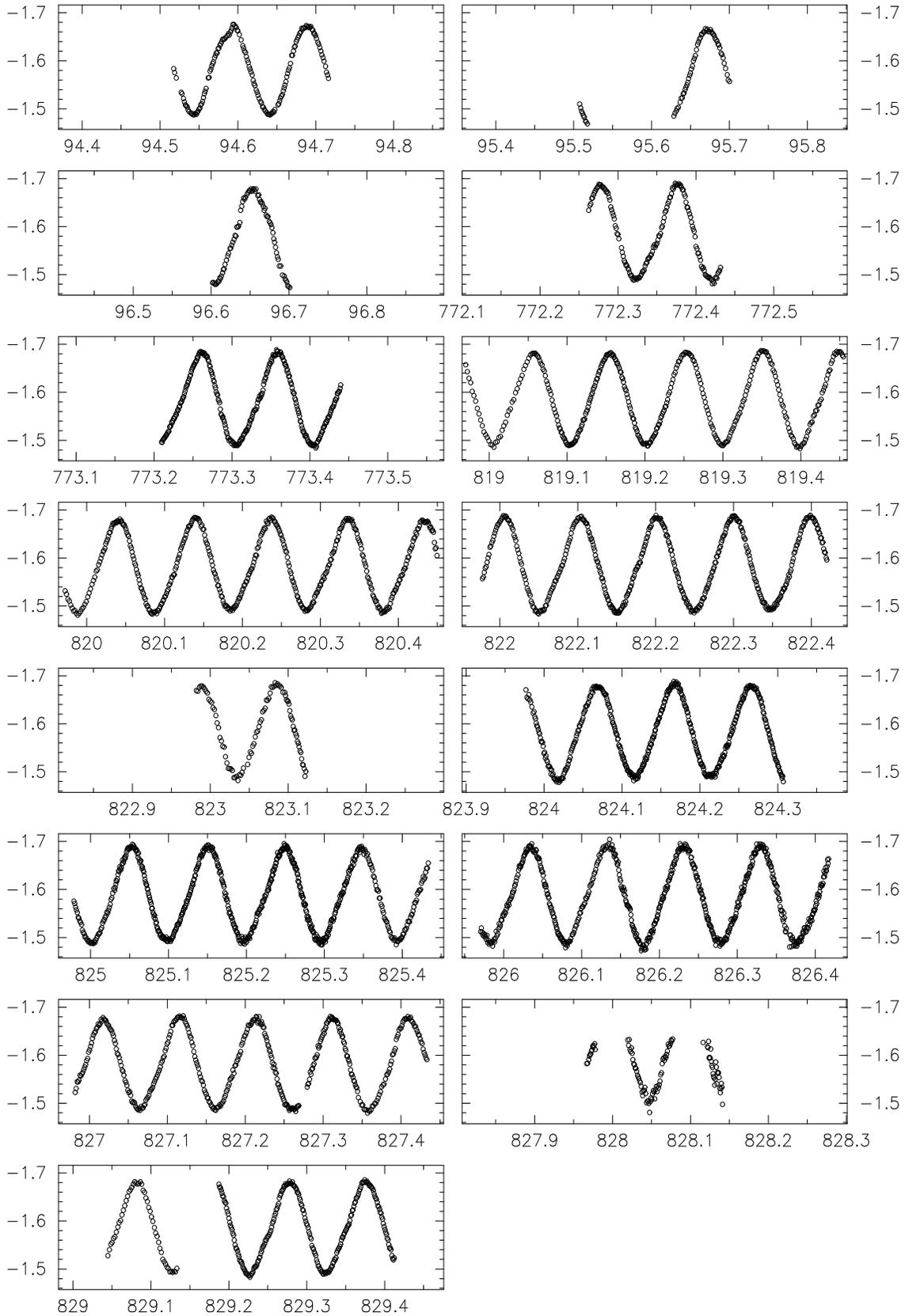}
  \caption{Light curves of AN Lyn: NAOC 2008--2010 in $V$ filter. Time in HJD 2454400+ (days). }
\label{Fig:LC2008-2010NAOC}
\end{figure}

\clearpage
\begin{figure}[tp!]
  \vspace{-5mm}
\includegraphics[width=115mm,height=145mm,angle=-90,clip=true]{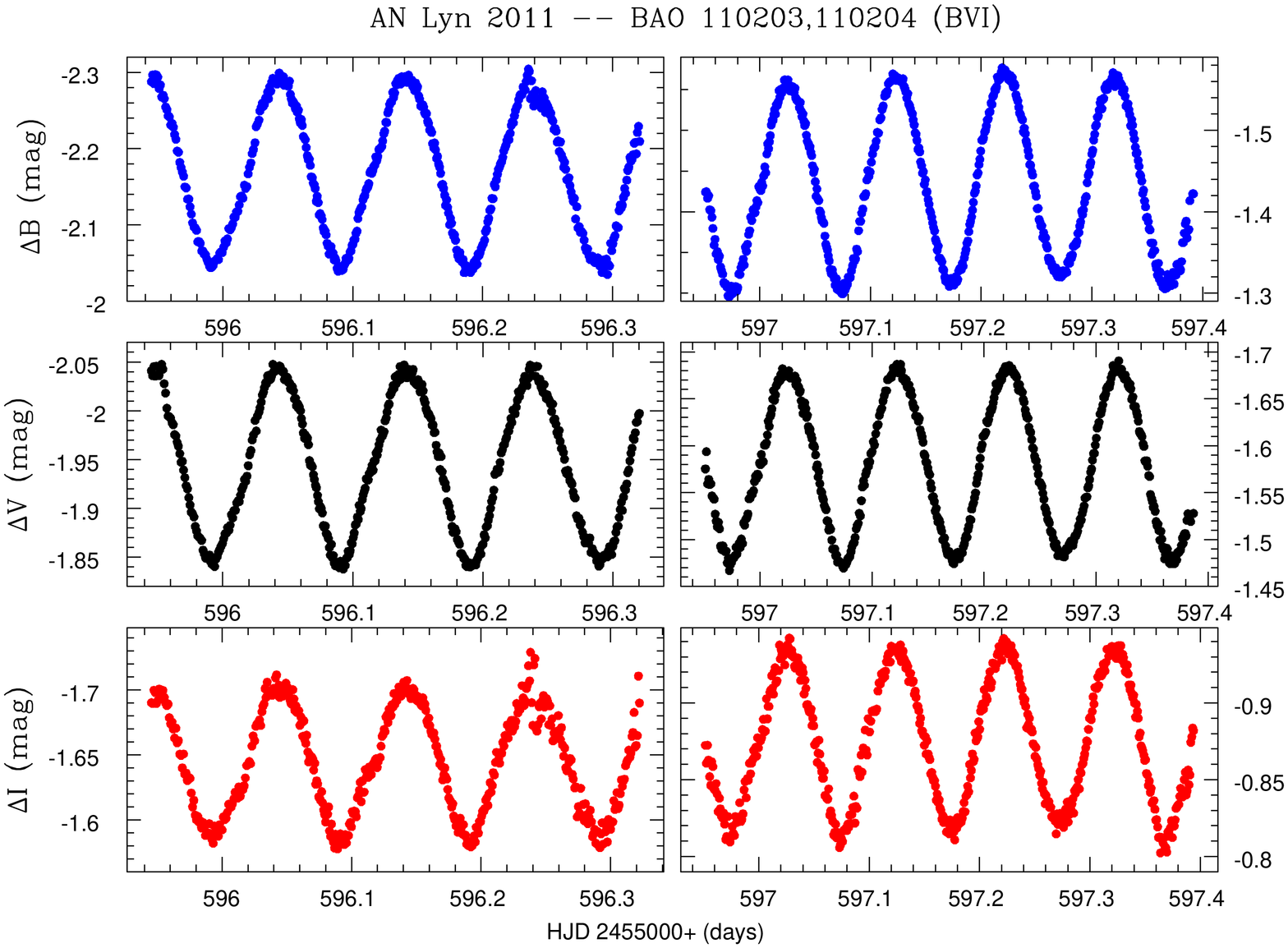}
\includegraphics[width=115mm,height=145mm,angle=-90,clip=true]{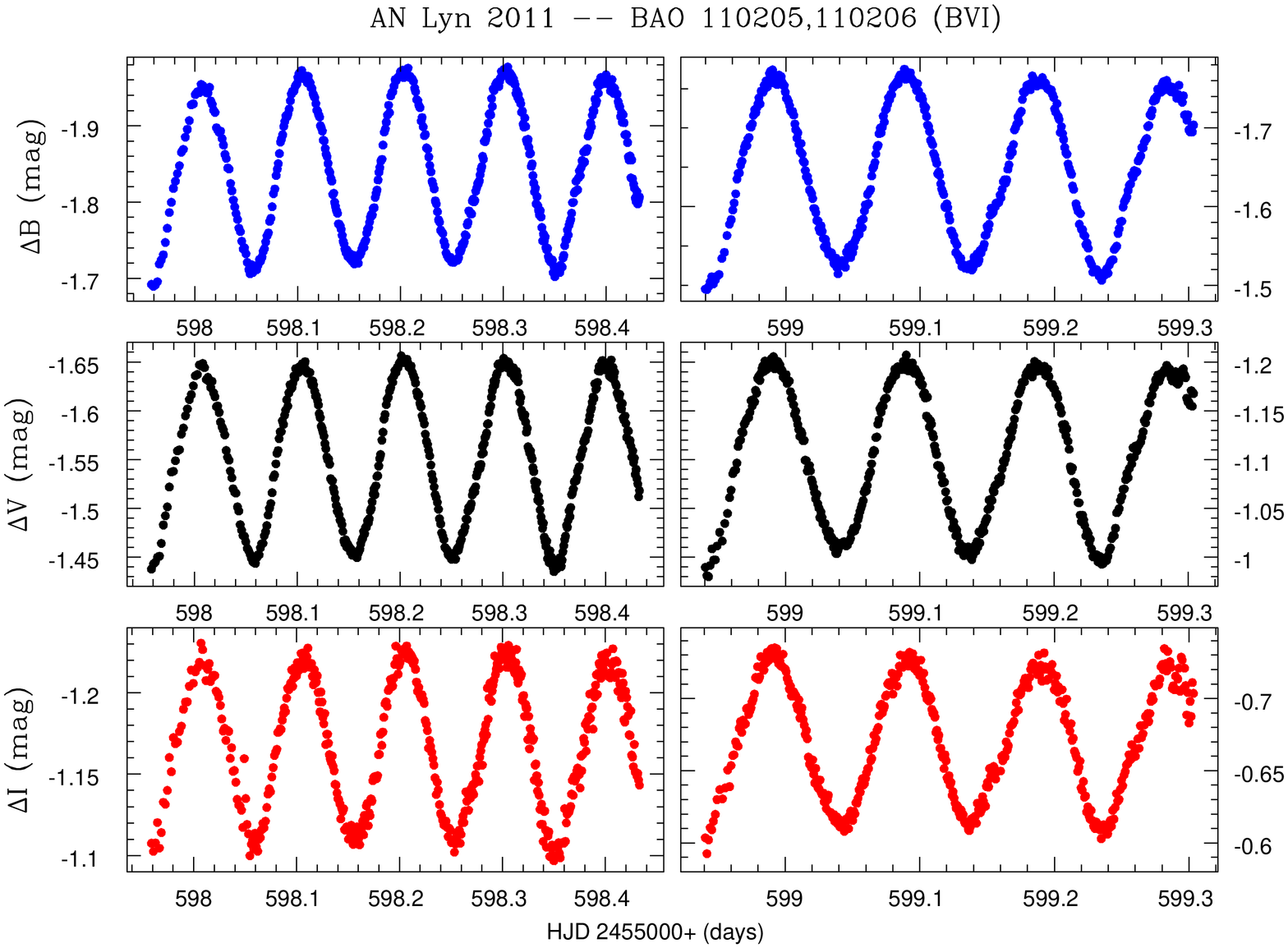}
  \caption{Light curves of AN Lyn: NAOC 2011 Feb 03,04,05,06 in $BVI$ filters. }
\label{Fig:LC2011NAOC}
\end{figure}

\clearpage
\addtocounter{figure}{-1}
\begin{figure}[tp!]
  \vspace{-5mm}
\includegraphics[width=115mm,height=145mm,angle=-90,clip=true]{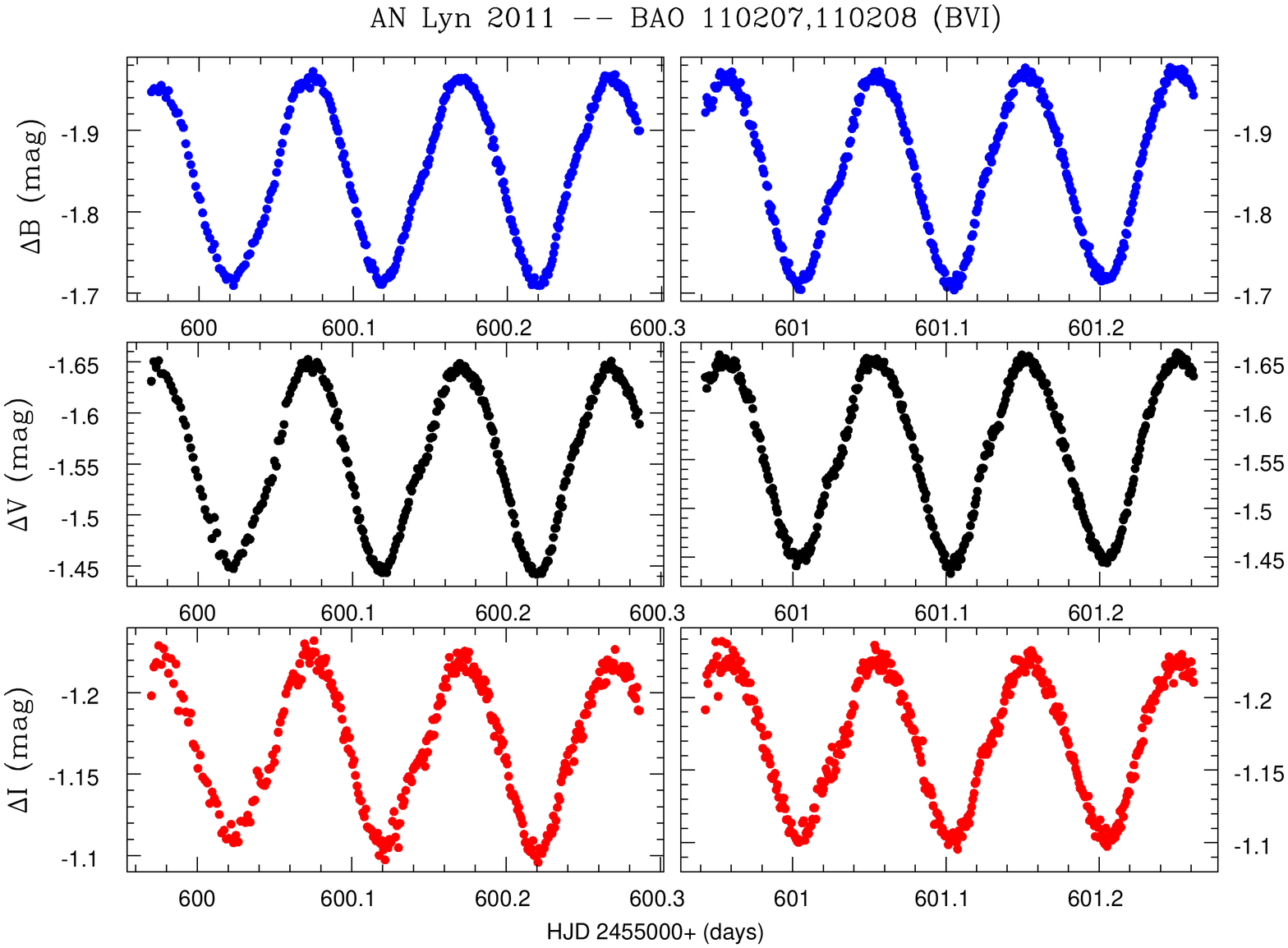}
\includegraphics[width=115mm,height=145mm,angle=-90,clip=true]{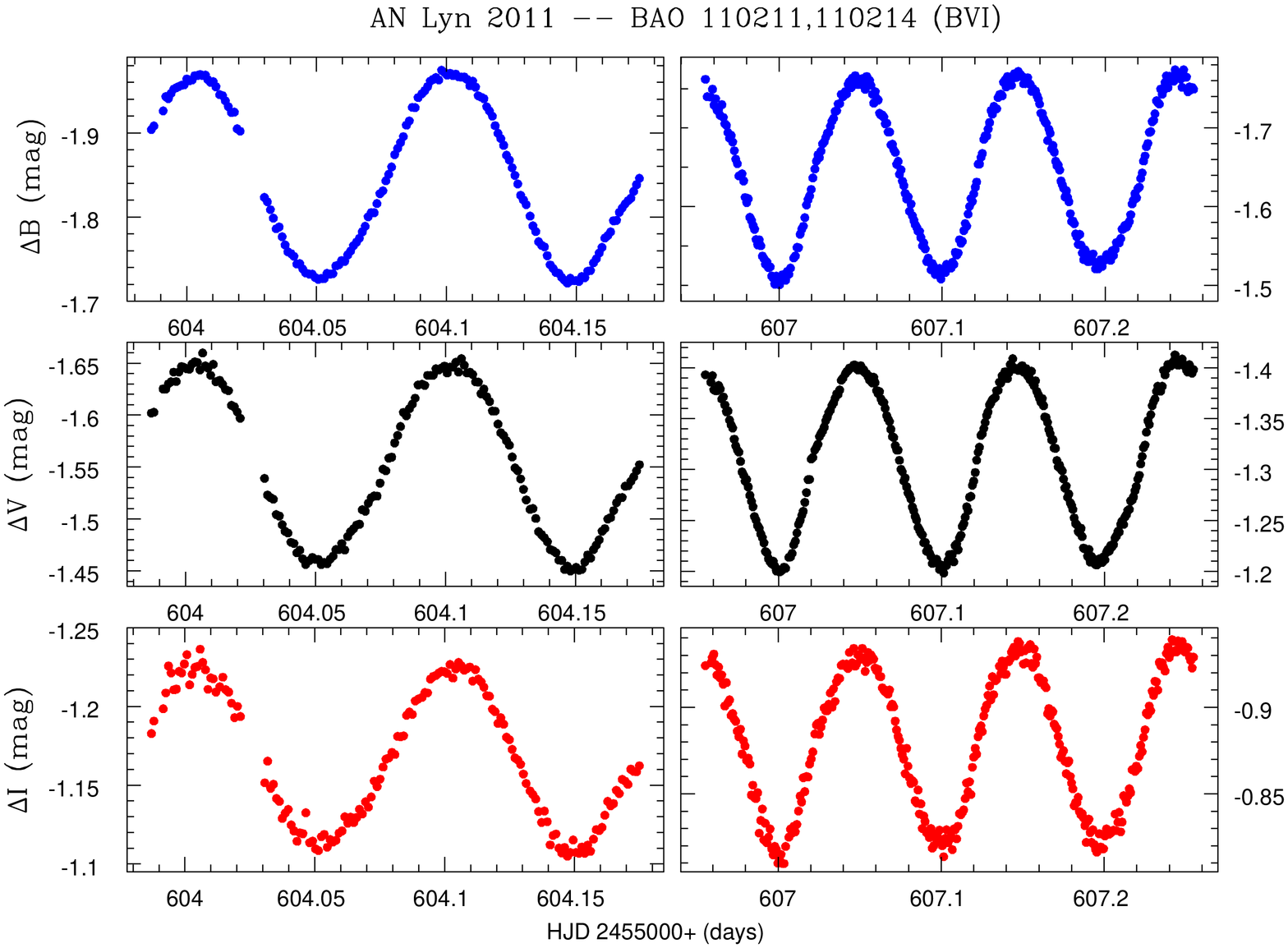}
  \caption{Light curves of AN Lyn: NAOC 2011 Feb 07,08,11,14 in $BVI$ filters. }
\end{figure}

\clearpage
\begin{figure}[t]
  \vspace{-8mm}
\centering
\includegraphics[width=115mm,height=120mm,angle=-90,clip=true]{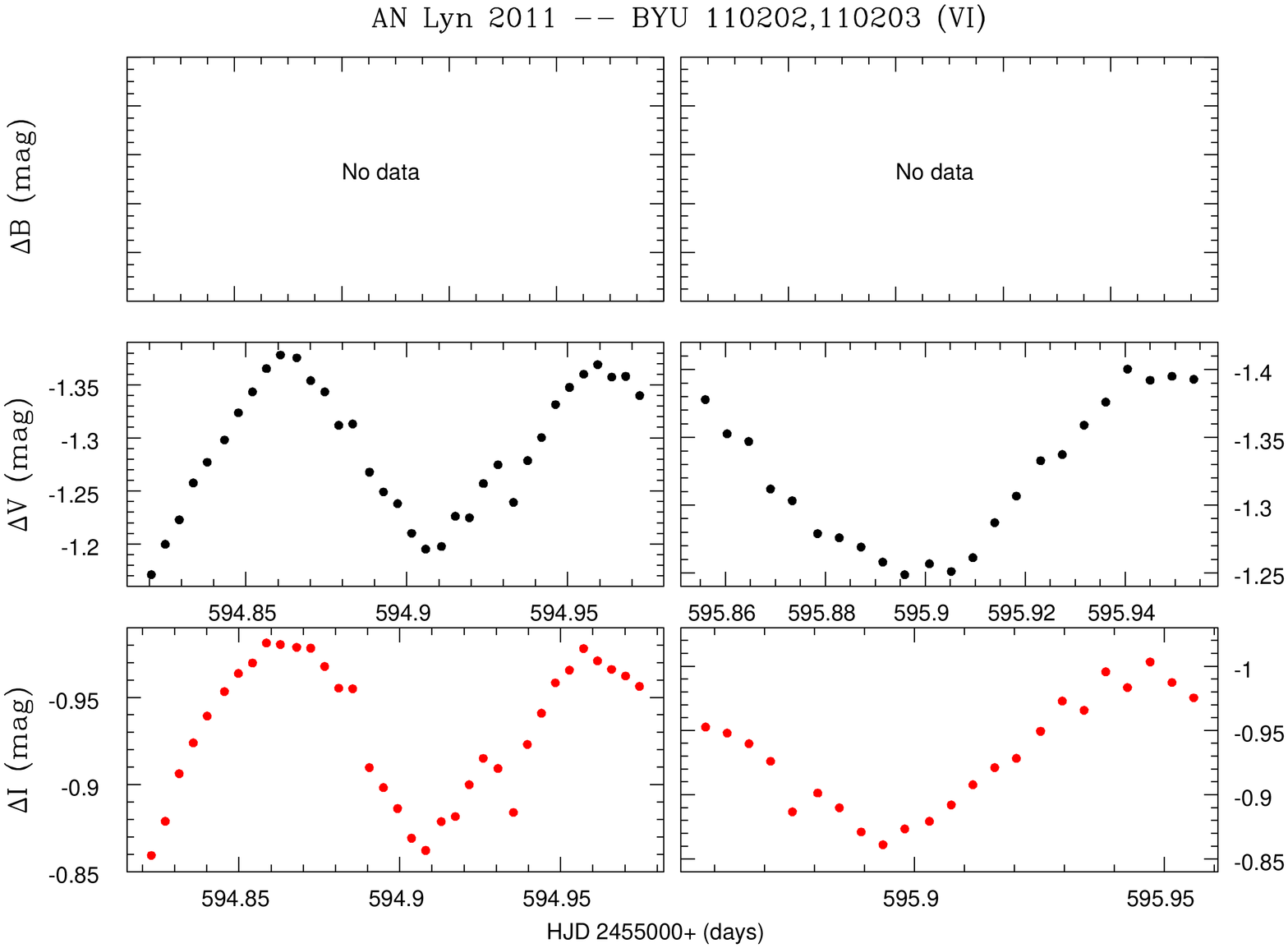}
\includegraphics[width=115mm,height=150mm,angle=-90,clip=true]{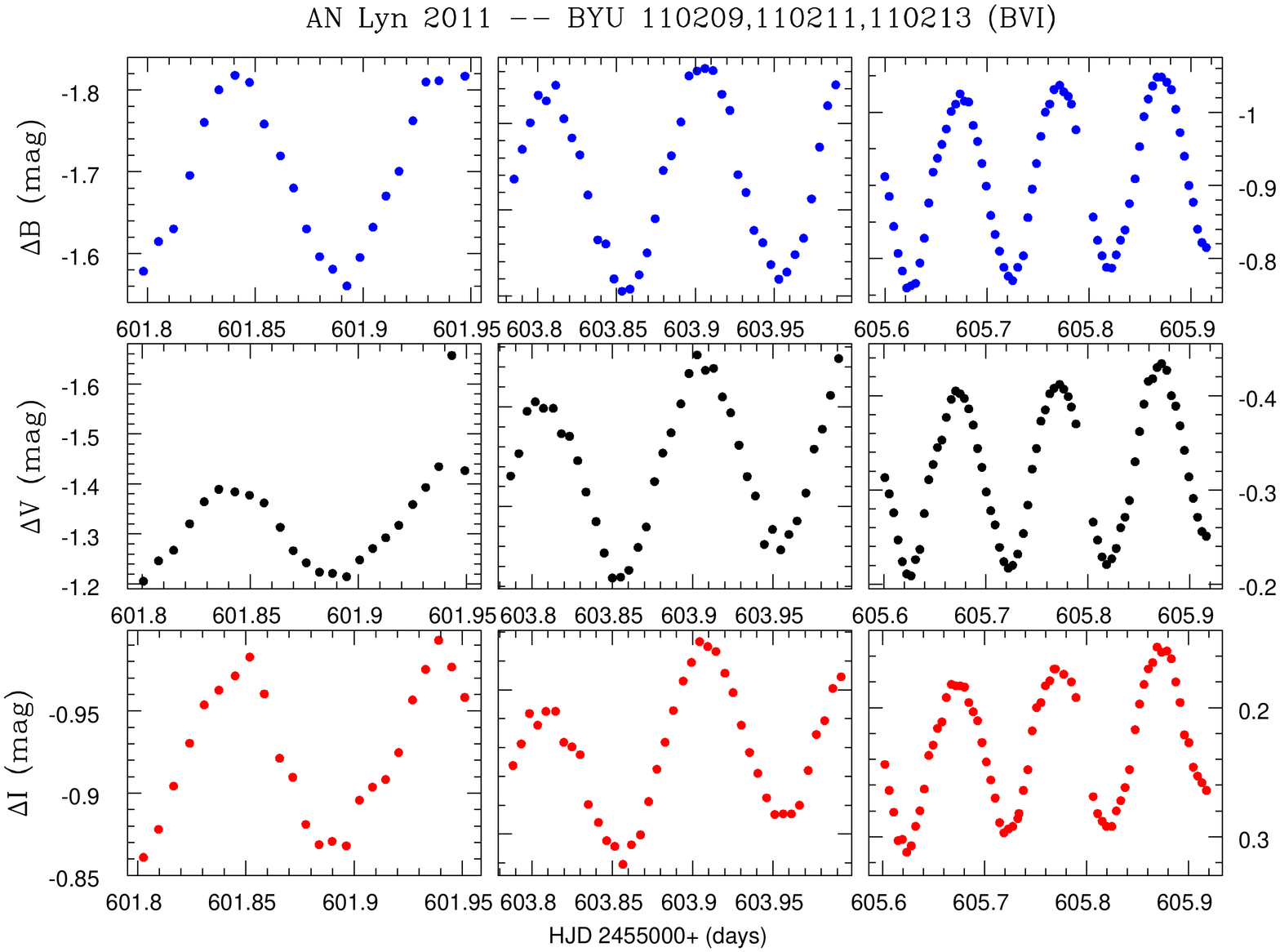}
  \caption{Light curves of AN Lyn: BYU 2011 Feb 02,03,09,11,13 in $BVI$ filters. }
\label{Fig:LC2011BYU}
\end{figure}

\begin{figure}[tp]
  \vspace{-8mm}
\includegraphics[width=120mm,height=150mm,angle=-90,clip=true]{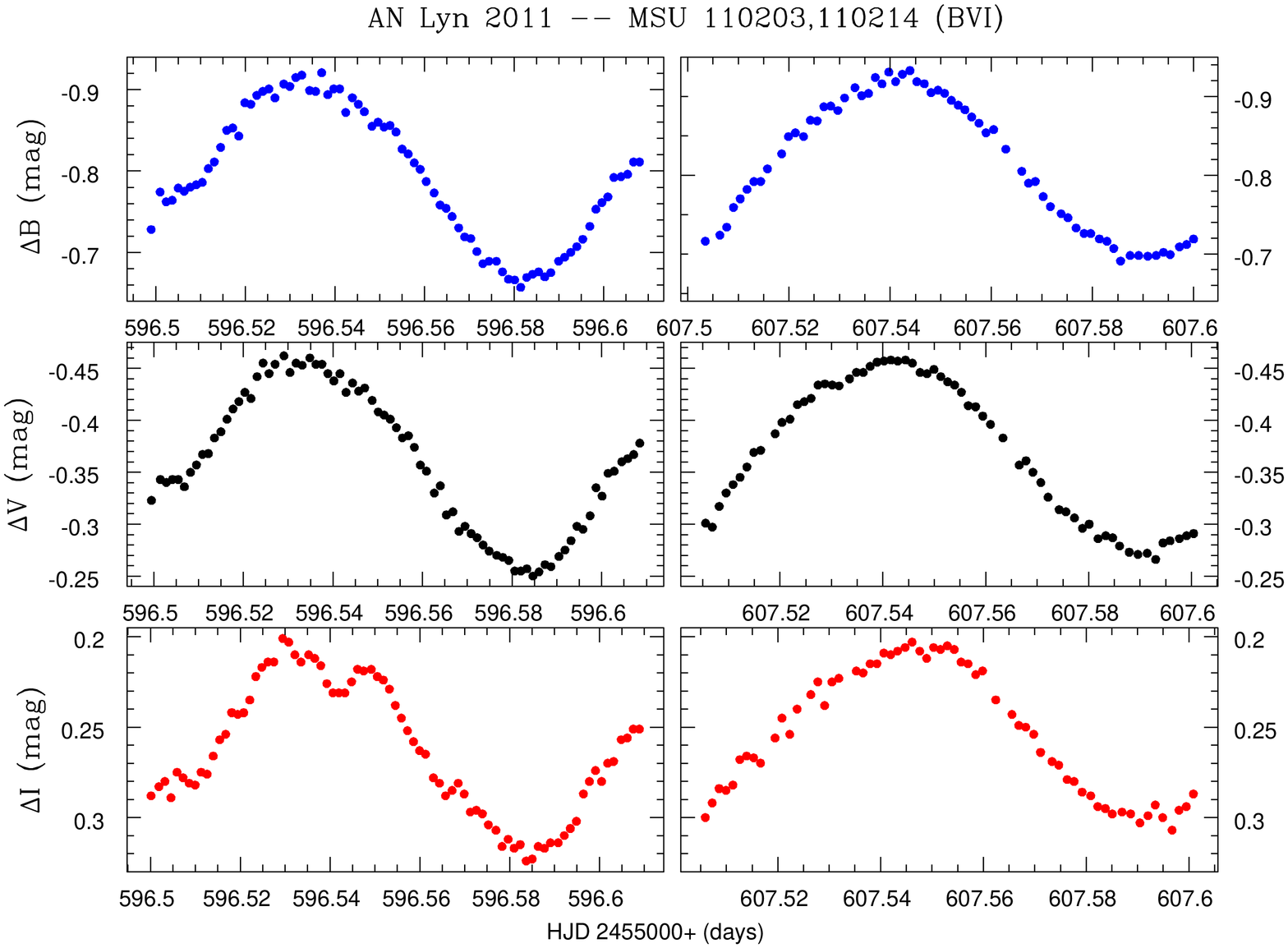}
  \caption{Light curves of AN Lyn: MSU 2011 Feb 03,14 in $BVI$ filters. }
\label{Fig:LC2011MSU}
\end{figure}

\clearpage
\begin{figure}[tp]
  \vspace{-8mm}
\includegraphics[width=115mm,height=150mm,angle=-90,clip=true]{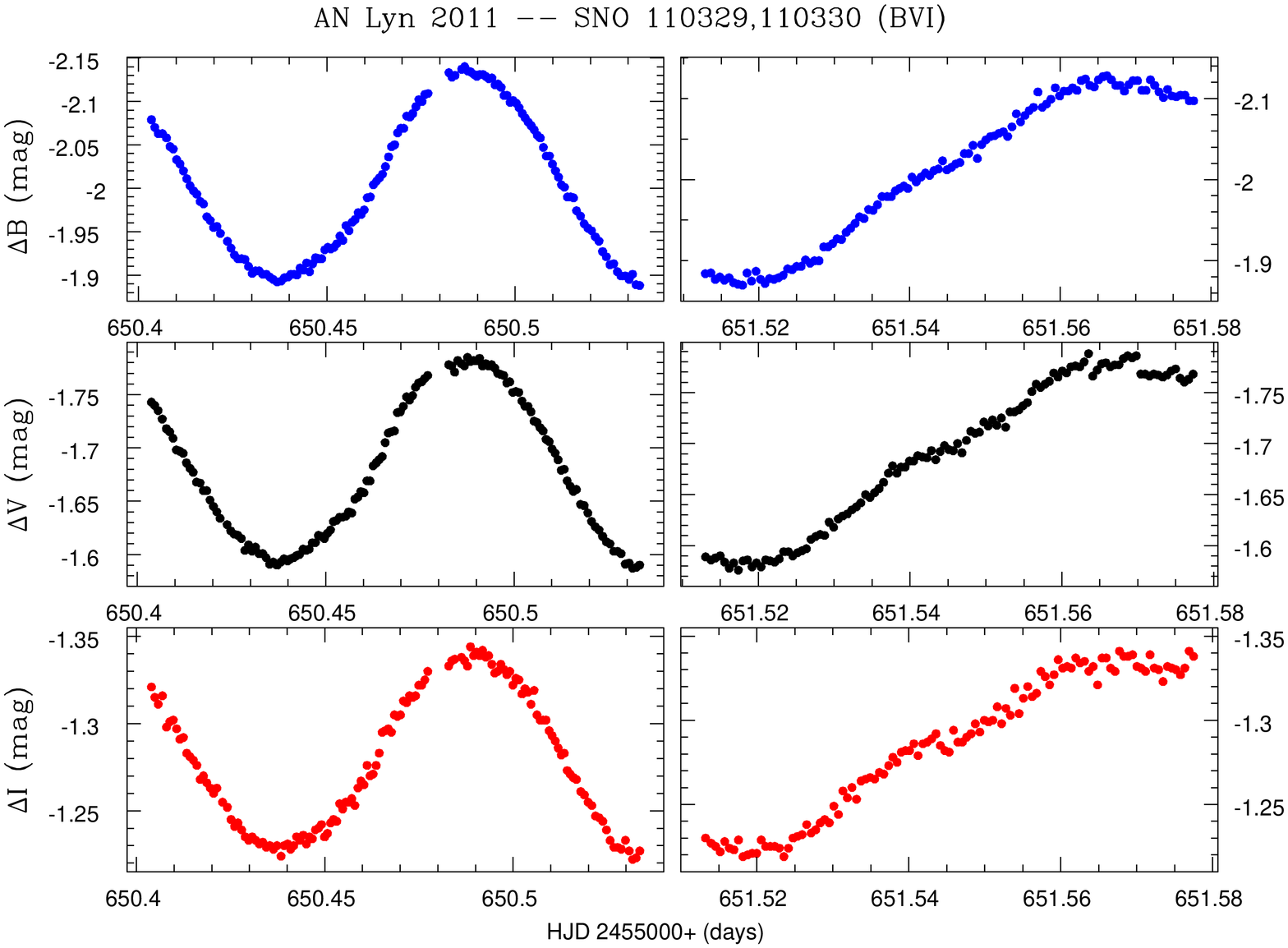}
\includegraphics[width=115mm,height=150mm,angle=-90,clip=true]{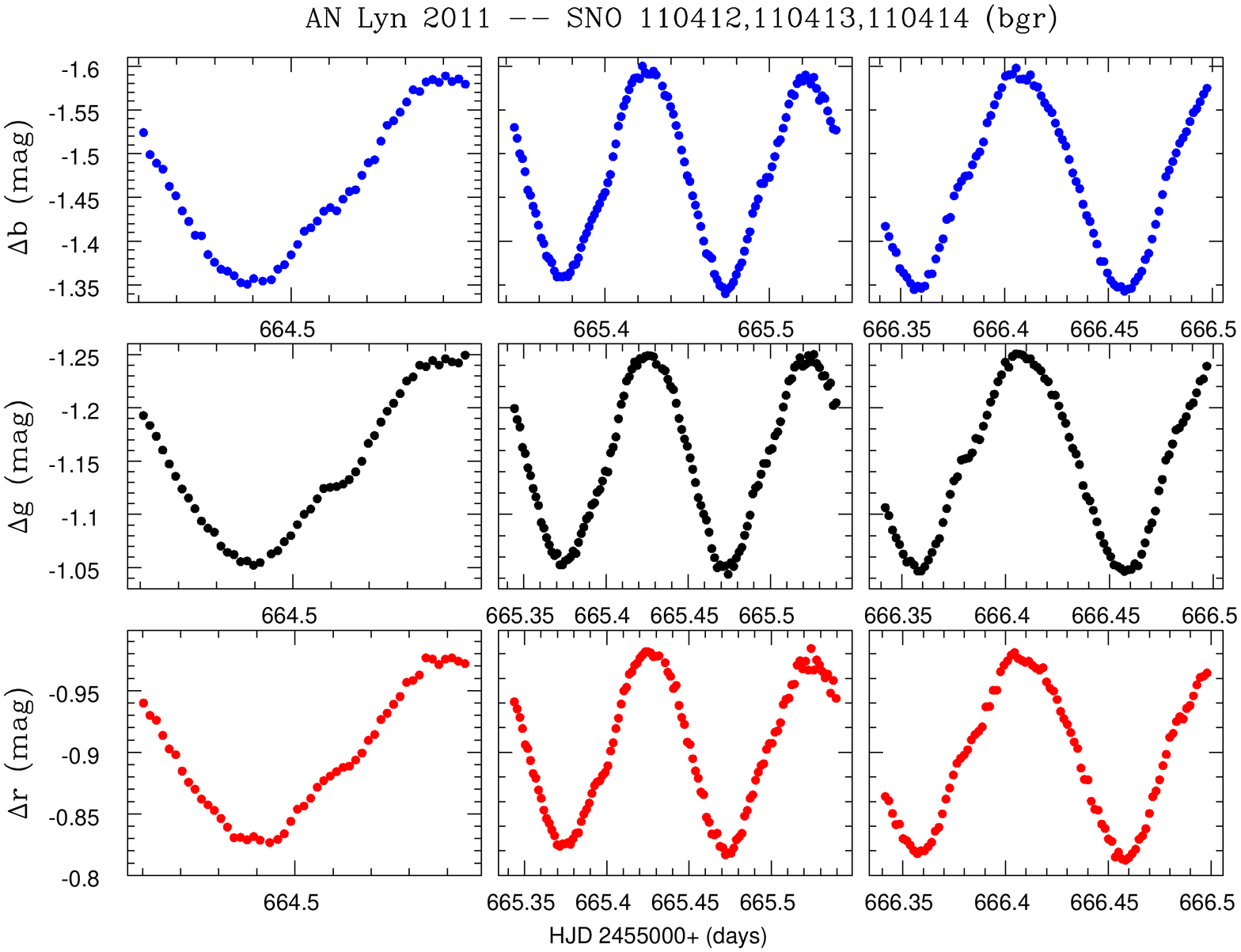}
  \vspace{-4mm}
  \caption{Light curves of AN Lyn: SNO 2011 Mar 29,30    in $BVI$ filters;
                                       2011 Apr 12,13,14 in $bgr$ filters. }
\label{Fig:LC2011SNO}
\end{figure}

\clearpage
\begin{figure}[t]
\vspace{-10mm}
\centering
\includegraphics[width=140mm,height=135mm,angle=-90,clip=true]{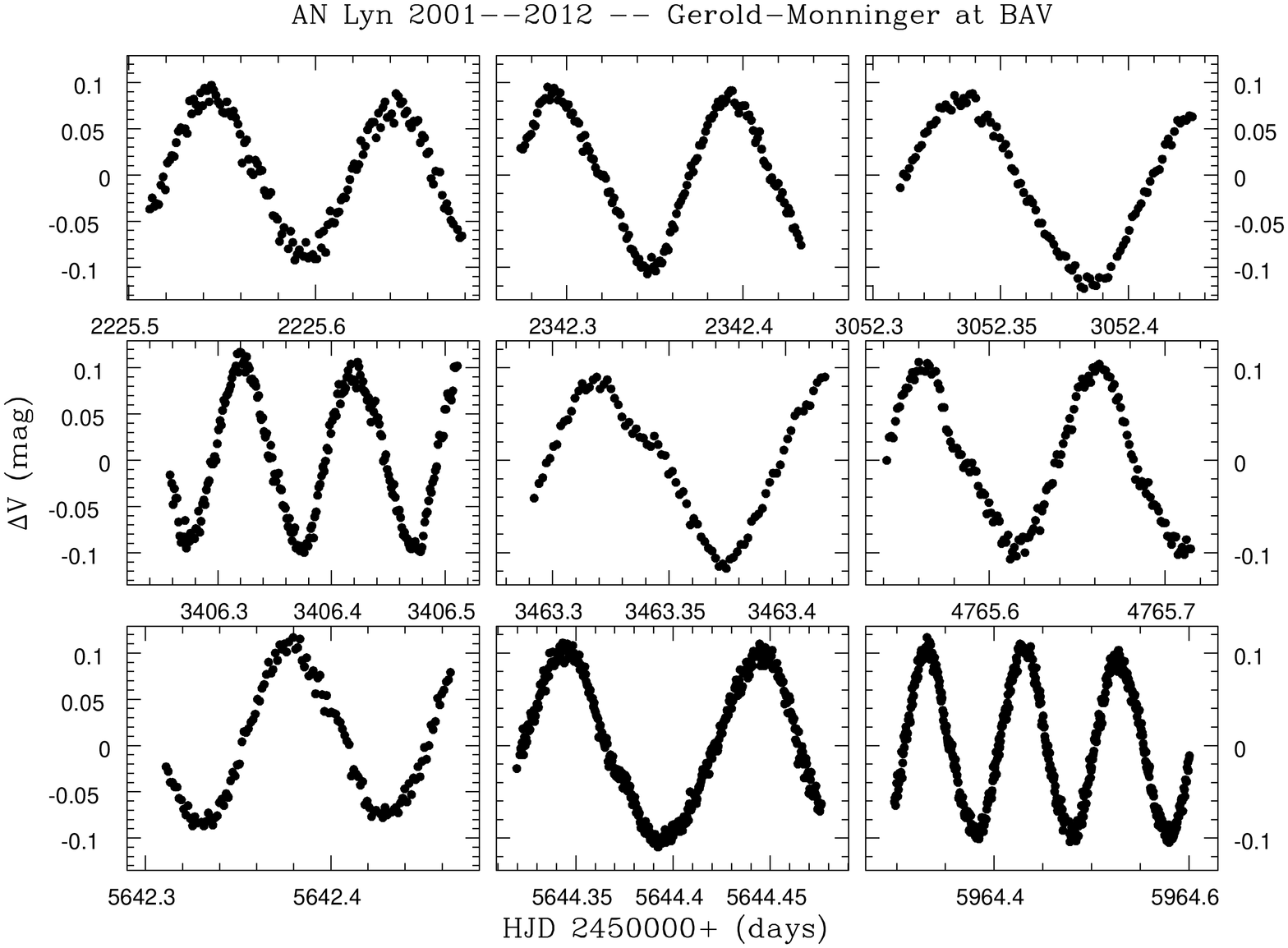}
\caption{Light curves of AN Lyn: BAV 2001--2012, 1841 points on 9 nights
in IR cut-off filter (2001) and $V$ filter (2002--2012). }
\label{Fig:LC2012BAV}
\end{figure}

\clearpage
\begin{figure}[t]
\vspace{-10mm}
\centering
\includegraphics[width=125mm,height=150mm,angle=-90,clip=true]{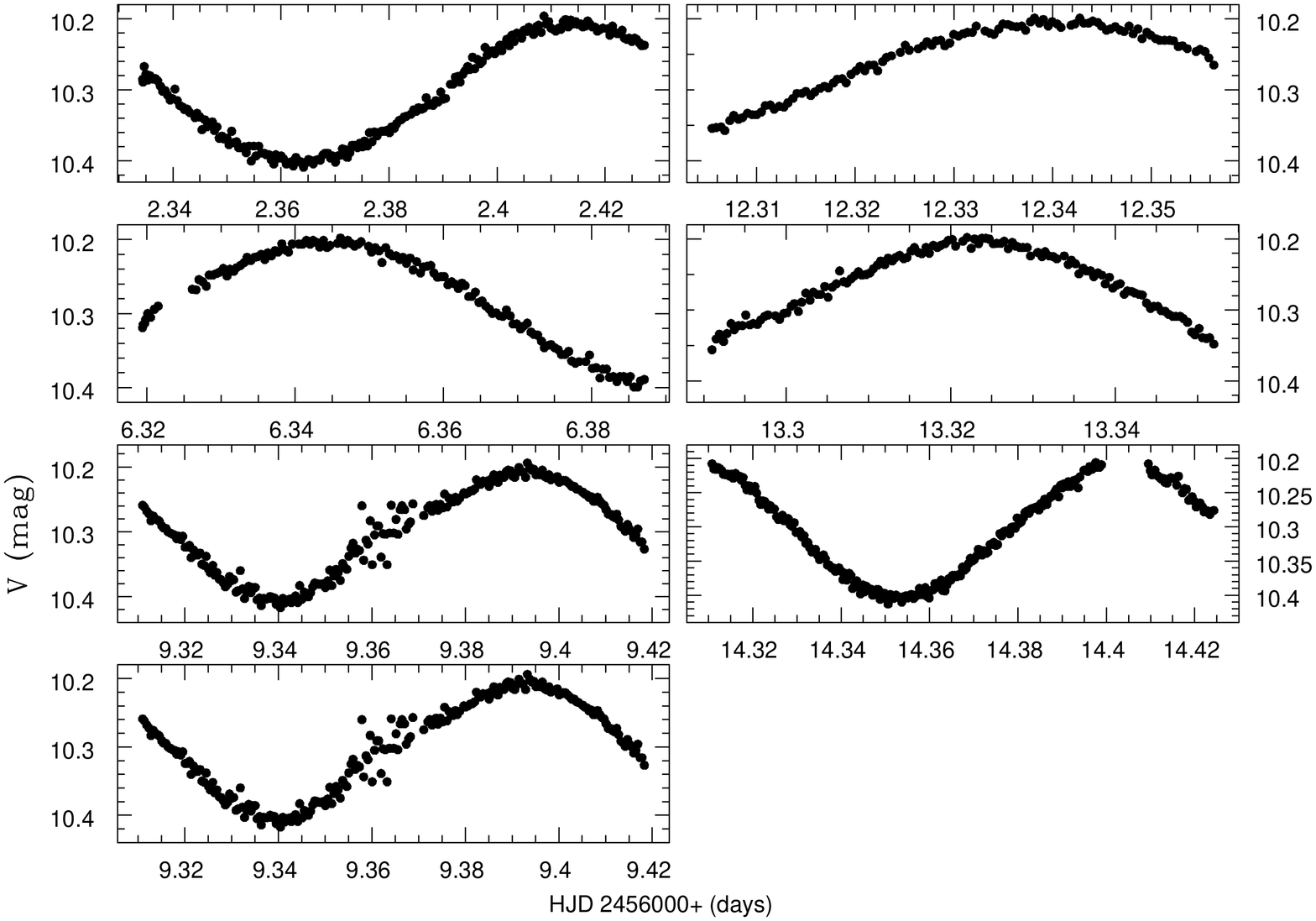}
\includegraphics[width=110mm,height=150mm,angle=-90,clip=true]{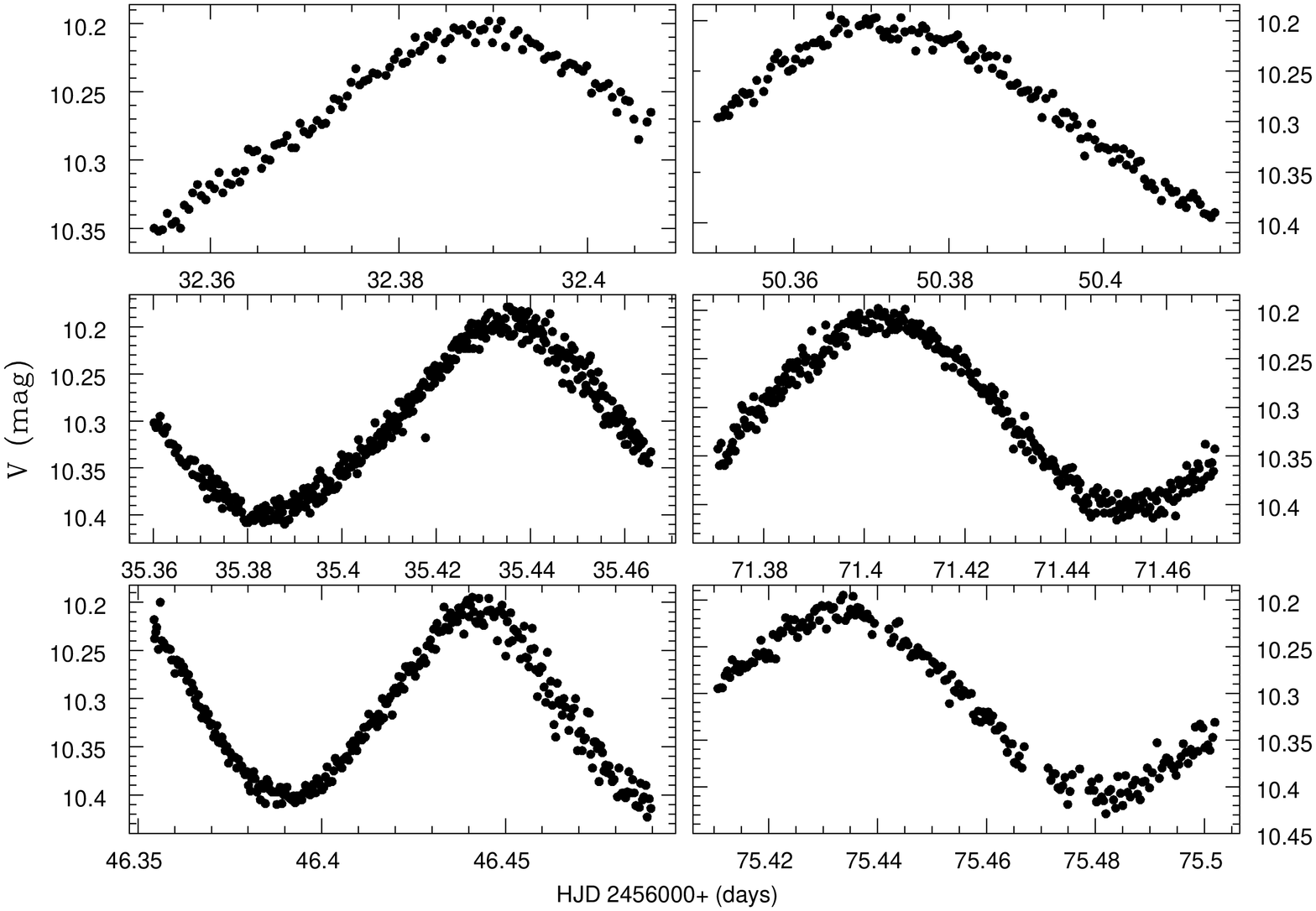}
\caption{Light curves of AN Lyn: Klockenhagen 2012, 13 nights in $V$ filter. }
\label{Fig:LC2012Pagel}
\end{figure}

\clearpage
\begin{table}[t]
  \caption[]{Observing log of photoelectric and CCD photometry of AN Lyn.
  Observing duration is given in HJD 2400000+ (days).
  The third column lists the numbers of either raw CCD frames or photoelectric records.
  $\Delta t$ is the exposure/integration time in seconds.
  `Instrumentation' column gives observatories' abbreviations, telescope sizes,
  photometers or CCD cameras, and filters. }
\end{table}
\addtocounter{table}{-1}
\begin{center}
\begin{longtable}{ccrccl}
  \label{Tab:Log}
\vspace{-13mm}
\\ \tableline
\multicolumn{1}{c} {UT Date} &
\multicolumn{1}{c} {Duration} &
\multicolumn{1}{r} {Frame} &
\multicolumn{1}{c} {Maximum} &
\multicolumn{1}{c} {$\Delta t$} &
\multicolumn{1}{l} {Instrumentation}  \\
\endfirsthead

\multicolumn{6}{c}%
{{\tablename\ \thetable{} -- continued from previous page}}
\vspace{3.0mm} \\
\tableline
\multicolumn{1}{c} {UT Date} &
\multicolumn{1}{c} {Duration} &
\multicolumn{1}{r} {Frame} &
\multicolumn{1}{c} {Maximum} &
\multicolumn{1}{c} {$\Delta t$} &
\multicolumn{1}{l} {Instrumentation}  \\
\tableline
\endhead

\tableline
\vspace{-2.0mm} \\
 \multicolumn{6}{c}{{Continued on next page}} \\
\endfoot
\tableline
\endlastfoot

\tableline\noalign{\smallskip}
2001.11.11 & 52225.511--225.678 &  139 & 1 & 30 & BAV 35cm, SBIG ST-6,IR cut-off \\
2002.02.01 & 52306.977--307.410 & 3684 & 4 & 10 & NAOC 85cm, 3-ch, V        \\ 
2002.02.02 & 52307.967--308.406 & 3716 & 4 & 10 & NAOC 85cm, 3-ch, V        \\
2002.02.03 & 52308.962--309.263 & 2441 & 3 & 10 & NAOC 85cm, 3-ch, V        \\
2002.02.04 & 52309.979--310.392 & 3412 & 4 & 10 & NAOC 85cm, 3-ch, V        \\
           & 52310.392--310.668 &  102 & 2 &    & SNO 90cm, $uvby\beta$ 6-ch\\
2002.02.05 & 52310.991--311.399 & 3481 & 4 & 10 & NAOC 85cm, 3-ch, V        \\
2002.02.06 & 52311.977--312.384 & 3444 & 4 & 10 & NAOC 85cm, 3-ch, V        \\
\hline
\end{longtable}
\vspace{-16mm}
\noindent
{\flushleft
\hspace{-0mm} Note: \dag ~total frames in all filters;\\
\hspace{-3mm}       \ddag ~exposure time is given only for
the first filter in the sequence for reference.
}
\end{center}

%
\clearpage
\begin{table}[t]
  \caption[]{List of 306 new times of maximum light of AN Lyn determined with the data
  collected in 2001--2012 by the authors. }
\label{Tab:max306}
\end{table}
\addtocounter{table}{-1}
\begin{center}
\begin{longtable}{lllll}
\vspace{-20mm}
\\ \tableline
\multicolumn{1}{c} {HJD$_{\rm max}$} &
\multicolumn{1}{c} {HJD$_{\rm max}$} &
\multicolumn{1}{c} {HJD$_{\rm max}$} &
\multicolumn{1}{c} {HJD$_{\rm max}$} &
\multicolumn{1}{c} {HJD$_{\rm max}$}  \\
\endfirsthead

\multicolumn{5}{c}%
{{\tablename\ \thetable{} -- continued from previous page}}
\vspace{3.0mm} \\
\tableline
\multicolumn{1}{c} {HJD$_{\rm max}$} &
\multicolumn{1}{c} {HJD$_{\rm max}$} &
\multicolumn{1}{c} {HJD$_{\rm max}$} &
\multicolumn{1}{c} {HJD$_{\rm max}$} &
\multicolumn{1}{c} {HJD$_{\rm max}$}  \\
\tableline
\endhead

\tableline
\vspace{-2.0mm} \\
 \multicolumn{5}{c}{{Continued on next page}} \\
\endfoot
\tableline
\endlastfoot

\tableline\noalign{\smallskip}
2452225.5947  & 2452314.53520 & 2452329.27613  & 2454131.05206  & 2454141.07514 \\
2452307.06608 & 2452315.12387 & 2452329.37290  & 2454131.34585  & 2454141.17314 \\
2452307.16433 & 2452315.22290 & 2452329.47270  & 2454132.03490  & 2454141.27176 \\
2452307.26328 & 2452315.31972 & 2452330.35620  & 2454132.13193  & 2454141.36533 \\
2452307.36067 & 2452315.32085 & 2452330.45262  & 2454132.23118  & 2454142.05792 \\
2452308.04870 & 2452315.51722 & 2452332.02783  & 2454132.32824  & 2454142.15658 \\
2452308.14553 & 2452315.61397 & 2452332.13249  & 2454133.01583  & 2454142.25529 \\
2452308.24608 & 2452316.00821 & 2452332.22230  & 2454133.11396  & 2454142.35045 \\
\hline
\end{longtable}
\end{center}
\newpage

\clearpage
\begin{table}[t]
  \caption[]{List of 541 maxima analyzed in present work.
  Fractional cycles are listed for check that whether or not an integer cycle number
  is properly rounded off. $(O-C)$ is based on the ephemeris~
${\rm HJD_{max}} = 2452307.077419 + 0.098274455\times E$ ~(eq.~\ref{eq:new-epoch}). }
\end{table}
\begin{center}
\addtocounter{table}{-1}
\begin{longtable}{ccrccl}
\label{Tab:max541}
\vspace{-6mm}
\\ \tableline
\multicolumn{1}{c} {No.} &
\multicolumn{1}{c} {HJD$_{\rm max}$} &
\multicolumn{1}{l} {Fractional Cycles} &
\multicolumn{1}{l} {Rounded Cycles} &
\multicolumn{1}{l} {$(O-C)$} &
\multicolumn{1}{l} {Source/Obs. Run}\\
\endfirsthead

\multicolumn{6}{c}%
{{\tablename\ \thetable{} -- continued from previous page}}
\vspace{3.0mm} \\
\tableline
\multicolumn{1}{c} {No.} &
\multicolumn{1}{c} {HJD$_{\rm max}$} &
\multicolumn{1}{l} {Fractional Cycles} &
\multicolumn{1}{l} {Rounded Cycles} &
\multicolumn{1}{l} {$(O-C)$} &
\multicolumn{1}{l} {Source/Obs. Run}\\
\tableline
\endhead

\tableline
\vspace{-2.0mm} \\
 \multicolumn{6}{c}{{Continued on next page}} \\
\endfoot
\tableline
\endlastfoot

\tableline\noalign{\smallskip}
     1 &  2444291.03400 &   -81567.9254 &     -81568 &   0.007326 & Costa et al.(1984)\\
     2 &  2444291.12500 &   -81566.9995 &     -81567 &   0.000052 & Costa et al.(1984)\\
     3 &  2444291.22200 &   -81566.0124 &     -81566 &  -0.001222 & Costa et al.(1984)\\
     4 &  2444292.11100 &   -81556.9663 &     -81557 &   0.003307 & Costa et al.(1984)\\
     5 &  2444292.21100 &   -81555.9488 &     -81556 &   0.005033 & Costa et al.(1984)\\
     6 &  2444300.07500 &   -81475.9280 &     -81476 &   0.007077 & Costa et al.(1984)\\
     7 &  2444300.16500 &   -81475.0122 &     -81475 &  -0.001198 & Costa et al.(1984)\\
     8 &  2444300.96000 &   -81466.9226 &     -81467 &   0.007606 & Costa et al.(1984)\\
\hline
\end{longtable}
\vspace{-16mm}
\noindent
\end{center}
{\flushleft
\hspace{-0mm} References: \footnotesize{
      Agerer et al.(1983): IBVS, 2370;  
      Agerer \& H\"{u}bscher(2002): IBVS, 5296;
      Agerer \& H\"{u}bscher(2003): IBVS, 5485;
      Costa et al.(1984): \aaps, 57, 233;
      Hintz et al.(2005): \aj, 130, 2876;
      H\"{u}bscher(2005): IBVS, 5643;
      H\"{u}bscher et al.(2005): IBVS, 5657;
      H\"{u}bscher et al.(2006): IBVS, 5731;
      H\"{u}bscher \& Walter(2007): IBVS, 5761;
      Klingenberg et al.(2006): IBVS, 5701;
      Li et al.(2010): PASJ, 62, 987;
      Pensado(1983): Bol. Aatron. Obs. Madr., 11, 3;
      Rodr\'{\i}guez et al.(1997a): A\&A, 324, 959;
      Rodr\'{\i}guez et al.(1997b): A\&A, 328, 235;
      Wils et al.(2010): IBVS, 5928;
      Zhou(2002): A\&A, 385, 503.
}}

\clearpage
%
%
\begin{sidewaystable}
\footnotesize
   \caption[ ]{Pulsation amplitudes (mmag in $V,y$) of AN Lyn for the grouped data sets.
   '1-f' column: single-frequency;
   '[$f_1$]' column: $f_1$-fixed;
   8--12th columns: five frequencies fixed.
   Columns N: number of nights;
   Max: number of maxima;
   Len: data length in hours;
   $\sigma$: standard deviation of residuals for the multifrequency fitting.}
\label{Tab:ampl}
\vspace{-2mm}
\begin{center}
\begin{tabular}{lrrrrccccccccl}
   \hline\noalign{\smallskip}
Dataset   &N    &Points& Max & Len   &1-$f$&$[f_1]$&  $f_1$ & $2f_1 $ & $f_{2}$ & $f_1$+$f_2$ & $2f_1$+$f_2$ & $\sigma$ & Source\\
   \hline\noalign{\smallskip}
1980      & --- & ---  & --- & ---   & --- &---   & 79.0$\pm$3.0 & --- & --- & --- &---  & --- & \citet{rodr97a} \\
1982      & --- & ---  & --- & ---   & --- &---   & 89.0$\pm$3.0 & --- & --- & --- &---  & --- & \citet{rodr97a} \\ 
1983      & --- & ---  & --- & ---   & --- &---   & 91.0$\pm$4.0 & --- & --- & --- &---  & --- & \citet{rodr97a} \\
1994      &  1  &   70 &   3 &   5.9 & 67.2& 67.2 & 66.5$\pm$2.7 & 7.3 & 2.3 & 1.1 & 0.1 & 4.2 & \citet{rodr97a}\\ 
1995      & 11  &  395 &  13 &  44.0 & 67.5& 67.5 & 67.3$\pm$2.1 & 5.6 & 3.8 & 2.6 & 0.9 &10.3 & \citet{rodr97a}\\ 
1996      & 16  &  531 &  19 &  49.9 & 71.8& 71.8 & 71.9$\pm$1.5 & 7.1 & 3.0 & 2.9 & 1.7 & 8.1 & \citet{rodr97b}\\ 
2000NAOC   & 21  & 2937 &  53 & 128.8 & 80.6& 80.6 & 80.7$\pm$1.5 & 4.8 & 4.8 & 3.3 & 1.9 &13.6& \citet{zhou02}\\
2000BYU   &  4  &  266 &   7 &  17.9 & 85.4& 85.2 & 84.8$\pm$2.7 & 5.8 & 4.5 & 5.7 & 2.6 & 9.4 & BYU unpublished\\
2001MSU   &  9  &  738 &   5 &  14.4 & 92.4& 90.2 & 89.3$\pm$2.7 &14.1 & 4.3 & 3.9 & 1.9 &15.9 & \citet{lac01} \\
2002MSU   &  4  &  116 &   2 &   4.6 & 94.5& 94.5 & 90.2$\pm$2.7 & 8.7 &10.5 & 5.8 & 2.2 & 9.0 & present work\\
2002SNO   & 12  &  846 &  18 &  47.5 & 88.6& 88.7 & 88.6$\pm$1.8 & 6.9 & 0.9 & 1.2 & 0.9 &10.9 & present work\\ 
2002NAOC  & 17  & 3989 &  57 & 149.9 & 89.0& 89.0 & 88.9$\pm$1.5 & 9.2 & 1.1 & 1.0 & 1.1 &18.2 & present work\\ 
2002NAOCD &  4  &  371 &   5 &  12.2 & 89.8& 89.8 & 89.9$\pm$1.8 & 4.8 & 2.9 & 2.0 & 6.2 & 8.5 & present work\\ 
2002      & 37  & 5322 &     & 214.2 & 90.0& 89.9 & 89.9$\pm$0.9 &10.3 & 1.4 & 0.8 & 1.3 &23.0 & present work\\
2005BYU   &  5  &  258 &   5 &  11.8 & 86.5& 86.4 & 86.3$\pm$2.7 & 9.0 & 9.6 & 4.6 & 0.5 &16.1 & BYU unpublished\\
2005M     &  2  &  296 &   4 &   9.1 & 92.7& 92.7 & 91.5$\pm$0.9 & 9.6 & 4.7 & 4.0 & 2.0 &11.1 & present work (Monninger)\\
2006BAK   &  3  &  276 &   4 &   9.9 & 91.3& 91.3 & 92.1$\pm$2.4 & 8.3 & 2.4 & 2.7 & 3.7 &11.2 & present work\\
2006BYU   &  2  &  348 &   4 &   9.3 & 91.7& 91.7 & 91.7$\pm$2.7 & 4.6 & 0.5 & 2.4 & 0.8 & 6.0 & BYU unpublished\\
2006K     &  1  &  155 &   2 &   5.3 & 89.4& 89.4 & 89.5$\pm$0.7 & 3.1 & 2.5 & 3.4 & 3.4 & 5.7 & Klingenberg et al. (2006)\\
2007BYU   &  3  &  434 &   4 &   8.7 & 90.9& 90.9 & 90.5$\pm$1.8 & 6.3 & 3.4 & 3.1 & 1.4 & 5.4 & BYU unpublished\\
2007NAOC  & 43  & 5137 & 104 & 275.4 & 89.0& 89.0 & 88.9$\pm$0.9 & 4.5 & 2.9 & 2.0 & 1.3 &12.8 & present work\\
2008NAOC  &  3  &  294 &   4 &  11.8 & 93.6& 93.8 & 93.8$\pm$2.7 & 2.6 & 1.2 & 3.4 & 2.2 & 5.9 & present work\\
2008BYU   &  5  &  303 &   3 &   9.0 & 94.9& 94.9 & 95.1$\pm$2.7 & 5.3 & 4.9 & 1.7 & 3.4 & 7.1 & BYU unpublished\\
2008Li    &  3  & 1181 &   4 &   7.8 & 95.0& 95.0 & 92.3$\pm$0.3 & 7.6 & 4.9 & 2.6 & 2.5 & 7.1 & Li et al. (2010)   \\
2009BYU   & 17  & 1638 &  12 &  43.0 & 93.4& 93.4 & 93.3$\pm$1.6 & 4.1 & 2.4 & 2.1 & 1.5 & 8.5 & BYU+MSU unpublished\\
2009NAOC  &  2  &  410 &   4 &   9.6 & 93.4& 92.9 & 92.9$\pm$2.4 &11.5 & 1.5 & 1.4 & 0.5 & 4.0 & present work\\
2009WC1   &  3  &  876 &   8 &  19.7 & 90.9& 90.9 & 90.7$\pm$0.4 & 4.1 & 3.3 & 2.4 & 1.7 & 8.5 & Wils et al. (2010), C1\\
2009WC2   & 11  & 1989 &  22 &  50.1 & 95.2& 95.2 & 95.1$\pm$0.7 & 3.8 & 4.2 & 2.4 & 3.0 &22.9 & Wils et al. (2010), C2\\
2010NAOC  & 10  & 3623 &  35 &  90.5 & 93.9& 93.9 & 93.7$\pm$1.5 & 6.5 & 2.2 & 1.4 & 0.7 & 7.6 & present work\\
2010BYU   & 12  &  329 &  10 &  26.8 & 89.4& 89.9 & 89.4$\pm$1.5 & 4.3 & 2.5 & 0.7 & 1.2 &15.9 & BYU unpublished\\
2009--2010& 12  & 4033 &  39 & 100.1 & 93.8& 93.8 & 93.6$\pm$1.5 & 6.9 & 2.3 & 1.1 & 0.7 & 7.6 & present work\\
2011      & 15  & 3945 &  46 & 126.8 & 93.6& 93.6 & 93.7$\pm$1.1 & 2.6 & 4.1 & 3.4 & 2.4 &19.9 & present work\\
2011M     &  2  &  509 &   3 &   7.5 & 95.2& 95.2 & 93.8$\pm$0.8 & 4.6 & 6.0 & 5.2 & 2.5 &12.1 & present work (Monninger)\\
2012M     &  1  &  524 &   3 &   7.3 & 96.7& 96.7 & 96.1$\pm$1.3 & 5.5 & 5.3 & 2.7 & 1.3 & 6.9 & present work (Monninger)\\
2012P1    &  7  & 1230 &   7 &  14.0 & 96.4& 96.4 & 95.6$\pm$0.3 & 5.6 & 3.3 & 2.7 & 0.8 & 6.9 & present work (Pagel)    \\
2012P     & 13  & 2702 &  13 &  26.0 & 96.6& 96.5 & 96.3$\pm$0.3 & 3.5 & 0.8 & 1.4 & 0.6 & 9.9 & present work (Pagel)    \\
   \hline\noalign{\smallskip}
\end{tabular}
\end{center}
\end{sidewaystable}
\normalsize
\clearpage

\end{document}